\documentclass[aps,prl,twocolumn,groupedaddress,nofootinbib,longbibliography]{revtex4-2}
\usepackage{amsmath, amsfonts, amssymb,amsthm}
\usepackage{graphicx,subfigure}
\usepackage{mathrsfs}
\usepackage[mathscr]{eucal}
\usepackage{mathtools}
\usepackage[T1]{fontenc}
\usepackage{physics}
\usepackage{bm}
\usepackage{subfigure}
\usepackage[bookmarks=false]{hyperref}
\usepackage{xcolor}
\usepackage{url}
\hypersetup{colorlinks=true,citecolor=blue,
linkcolor=blue,urlcolor=blue,pdfstartview=FitH,
bookmarksopen=true}

\newtheorem{definition}{Definition}

\usepackage[skins,breakable]{tcolorbox}
\tcbuselibrary{theorems}
\newtcbtheorem[number within=section, use counter*=thm]{mythm}{Theorem}%
{breakable,colback=red!5,colframe=red!35!red,fonttitle=\bfseries}{thm}
\newtcbtheorem[number within=section, use counter*=thm]{mylem}{Lemma}%
{breakable,colback=green!5,colframe=green!35!black,fonttitle=\bfseries}{lem}
\newtcbtheorem[number within=section, use counter*=thm]{mydef}{Definition}%
{breakable,colback=yellow!5,colframe=yellow!70!black,fonttitle=\bfseries}{def}
\newtcbtheorem[number within=section, use counter*=thm]{mycor}{Corollary}%
{breakable,colback=blue!5,colframe=blue!70!black,fonttitle=\bfseries}{cor}
\newtcbtheorem[number within=section, use counter*=thm]{myobs}{Observation}%
{breakable,colback=green!5,colframe=green!70!black,fonttitle=\bfseries}{obs}
\newtcbtheorem[number within=section, use counter*=thm]{mypro}{Proposition}%
{breakable,colback=orange!5,colframe=orange!70!black,fonttitle=\bfseries}{pro}

\begin{document}
\title{Krylov shadow tomography: Efficient estimation of quantum Fisher information}
\author{Da-Jian Zhang}
\email{zdj@sdu.edu.cn}
\affiliation{Department of Physics, Shandong University, Jinan 250100, China}
\author{D.~M.~Tong}
\email{tdm@sdu.edu.cn}
\affiliation{Department of Physics, Shandong University, Jinan 250100, China}

\date{\today}

\begin{abstract}
  Efficiently estimating the quantum Fisher information (QFI) is pivotal in quantum information science but remains an outstanding challenge for large systems due to its high nonlinearity. In this Letter, we tackle this long-standing challenge by integrating the Krylov subspace method--a celebrated tool from applied mathematics--into the framework of shadow tomography. The integrated technique, dubbed Krylov shadow tomography (KST), enables us to formulate a strict hierarchy of non-polynomial lower bounds on the QFI, among which the highest one matches the QFI exactly. We show that all the bounds can be expressed as expected values of the inverses of Hankel matrices, which are accessible via shadow tomography. Our KST therefore opens up a resource-efficient and experimentally feasible avenue to estimate not only non-polynomial lower bounds but also the QFI itself.
\end{abstract}

\maketitle

Nearly half a century after its inception \cite{1976Helstrom}, the quantum Fisher information (QFI) has proven to be of universal importance in quantum information science \cite{Toth2014JPAMT}. Initially introduced to describe the ultimate precision allowed by quantum mechanics \cite{1976Helstrom,Toth2014JPAMT,1994Braunstein3439,Braunstein1996AP,2017Seveso12111,2020Zhang23418,Zhang2022nQI,Zhou2024PRR}, the QFI has become the cornerstone of quantum metrology and underpinned vast applications ranging from phase estimation \cite{2006Giovannetti10401} to gravitational wave detection \cite{Schnabel2010NC} and atomic clock calibration \cite{Roslund2024N}. Its impact has extended beyond metrology and permeated into various other areas. The close relation between the QFI and entanglement, for instance, has inspired a variety of QFI-based criteria for entanglement detection \cite{Boixo2008PRL,Pezze2009PRL,Hyllus2012PRA,Toth2012PRA,Toth2018PRL,Toth2020PRL,Pal2021PRR,Ren2021PRL,Yadin2021NC,Tan2021PRL,Gianani2022PRA,Yang2024PRL,2024Imai,Szalay}. Now, with the advent of noisy intermediate-scale quantum (NISQ) era \cite{Preskill2018}, the relevance of the QFI expands even further, owing to newly discovered applications in emerging fields such as variational quantum algorithms and quantum machine learning \cite{Meyer2021Q,Jiao2023AQT}.

These widespread applications underscore the imperative need for efficiently estimating the QFI, which has been a long-standing challenge for large systems \cite{Gebhart2023NRP} despite substantial efforts \cite{Strobel2014S,Barontini2015S,Hauke2016NP,Froewis2016PRL,Lu2020PRL,Yang2020nQI,Yu2022nQI}. A fundamental difficulty is that the QFI is highly nonlinear and cannot be expressed as a polynomial of the density matrix. This contrasts with the current research landscape, where state-of-the-art techniques like shadow tomography \cite{Elben2022NRP,Cieslinski2024PR} thrive on estimating polynomial quantities like expectation values of observables \cite{Huang2020}, purities \cite{Enk2012PRL,Elben2018PRL,Brydges2019S}, state overlaps \cite{Elben2020PRL}, and statistical moments \cite{Elben2020,Neven2021nQI,Yu2021PRL,Liu2022PRL}.  Consequently, numerous studies have shifted their focus from estimating the QFI itself to instead accessing polynomial lower bounds for it \cite{Rivas2008PRA,Rivas2010PRL,Zhang2017PRA,Apellaniz2017PRA,Totha,Gaerttner2018PRL,Cerezo2021QST,Yu2021PRR}. This shift is, however, a mixed blessing. On one hand, polynomial lower bounds are relatively easier to access due to their simplicity and can provide a range within which the true QFI lies. On the other hand, a substantial gap exists between the bounds and the QFI such that QFI-based applications could be severely compromised in effectiveness if the bounds are used as substitutes. Recently, it has been proposed to reduce the gap by employing an infinite number of polynomial lower bounds that converge to the QFI asymptotically  \cite{Rath2021PRL,Vitale2024PQ}. While this proposal is theoretically intriguing, the required number of measurements and postprocessing grows exponentially with the degree of polynomials and is thus difficult to afford in practice \cite{Zhou2024nQI,Peng-arXiv}.

In this Letter, we integrate the Krylov subspace method into the framework of shadow tomography, for breaking ground in the latter to efficiently estimate non-polynomial quantities like the QFI. Renowned in the field of applied mathematics \cite{Higham2015}, the Krylov subspace method was initially devised to solve high-dimensional linear algebra problems. Its immense utility in the quantum domain has been recognized recently \cite{Nandy2024}, facilitating advancements in condensed-matter and many-body physics, including characterization of operator growth \cite{Parker2019PRX,Chu2024PRL}, quantum simulation of many-body dynamics \cite{Motta2019NP}, and quantum control of many-body systems \cite{Takahashi2024PRX}. The integrated technique we develop, dubbed Krylov shadow tomography (KST), leverages the Krylov subspace method to construct a nested sequence of Krylov subspaces, within each of which we formulate an optimal approximation to the QFI. This results in a strict hierarchy of non-polynomial lower bounds on the QFI. The most striking feature of our formulation is that the highest bound in this hierarchy matches the QFI exactly. We show that all the bounds can be expressed as expected values of the inverses of Hankel matrices, which are accessible via shadow tomography. Accordingly, only single-qubit gates and local measurements are required in our KST. Applying the KST to two physical contexts, we demonstrate that it allows for estimating the QFI itself at a moderate cost of resources, which is a sought-after goal that cannot otherwise be reached using polynomial lower bounds. The KST therefore opens up a resource-efficient and experimentally feasible avenue to estimate not only non-polynomial lower bounds but also the QFI itself.

\textit{Background.---}To illustrate the significance of the QFI, it is instructive to revisit its applications in entanglement detection. Let $\rho$ be an unknown state of $N$ qubits. Its QFI with respect to a Hermitian operator $H$ is defined as
\begin{eqnarray}
    F_Q=2\sum_{p_k+p_l>0}\frac{(p_k-p_l)^2}{p_k+p_l}\abs{\bra{k}H\ket{l}}^2,
\end{eqnarray}
where $p_k$ and $\ket{k}$ are the eigenvalues and eigenvectors of $\rho$. It has been shown that the inequality $F_Q\leq N$ holds for all separable states when $H=\frac{1}{2}\sum_{i=1}^N\sigma_z^{(i)}$ \cite{Pezze2009PRL}, with $\sigma_z^{(i)}$ denoting the Pauli-$Z$ matrix for qubit $i$. So a violation of this inequality, $F_Q>N$, serves as a criterion certifying that $\rho$ is entangled. 
This insight has inspired a series of studies that refine and extend the criterion to account for states of given entanglement depth \cite{Toth2012PRA,Hyllus2012PRA}, states characterized by Young diagrams \cite{Ren2021PRL}, networked states in distributed sensing \cite{Yang2024PRL}, and situations where $H$ is nonlinear \cite{2024Imai}. Apparently, practical implementations of these criteria are hindered by the difficulty in accessing $F_Q$. To address this issue, it has been suggested to alternatively estimate lower bounds $B$ that satisfy $B\leq F_Q$ for all states. The underlying rationale is that, to certify that $F_Q>N$ and thus $\rho$ is entangled, it suffices to show $B>N$. A simple lower bound, derived from the Legendre transform of $F_Q$, is proposed in Ref.~\cite{Apellaniz2017PRA}, given by
\begin{eqnarray}
    B^{(\mathsf{Leg})}=
    \begin{cases}
        N^2(1-2f_\textrm{GHZ})^2 & \textrm{if}~~f_\textrm{GHZ}>\frac{1}{2},    \\
        0                        & \textrm{if}~~f_\textrm{GHZ}\leq\frac{1}{2},
    \end{cases}
\end{eqnarray}
where $f_\textrm{GHZ}=\bra{\textrm{GHZ}}\rho\ket{\textrm{GHZ}}$ denotes the fidelity between $\rho$ and the Greenberger-Horne-Zeilinger
(GHZ) state. Another simple lower bound, grounded in the concept of the sub-QFI, is introduced in Ref.~\cite{Cerezo2021QST}, and reads
\begin{eqnarray}
    B^{(\mathsf{Sub})}=-2\tr([\rho, H]^2).
\end{eqnarray}
These bounds are polynomials of $\rho$ and easier to access in experiments. Yet, the substantial gap between them and $F_Q$ severely limits their effectiveness in detecting entanglement; that is, many entangled states detectable via $F_Q$ cannot be detected using these bounds. To reduce the gap, an infinite number of polynomial lower bounds, obtained from the Taylor expansion of $F_Q$, are proposed in Ref.~\cite{Rath2021PRL},
\begin{align}
    B_n^{(\mathsf{Tay})} &= 2\tr\Biggl( \sum_{l=0}^n \left(\rho \otimes \openone - \openone \otimes \rho \right)^2 
    \left( \openone \otimes \openone - \rho \otimes \openone \right. \nonumber\\
    &\quad - \left. \openone \otimes \rho \right)^l S(H \otimes H) \Biggr),
\end{align}
where $S$ is the swap operator. Remarkably, $B_n^{(\mathsf{Tay})}$ may increase as $n$ increases and eventually converges to $F_Q$ as $n\rightarrow \infty$. Unfortunately, it is very difficult, if not impossible, to obtain the QFI itself in this way, as the experimental effort in estimating $B_n^{(\mathsf{Tay})}$ grows exponentially with $n$.

\textit{Non-polynomial lower bounds in Krylov subspaces.---}To open the possibility of accessing the QFI, our idea is to employ the Krylov subspace method to formulate non-polynomial lower bounds that not only can effectively approximate the QFI but also be accessible via shadow tomography.

{First, we derive a compact expression for the QFI.} We define a weighted inner product between two Hermitian operators $X$ and $Y$ as $\langle X, Y\rangle _{\rho}\coloneqq\tr[\rho(XY+YX)/2]$, where the subscript $\rho$ is used to indicate the dependence of the inner product on $\rho$. The corresponding norm then reads $\norm{X}_\rho\coloneqq\sqrt{\langle X, X\rangle _{\rho}}$. Inspired by the seminal work of Braunstein and Caves \cite{1994Braunstein3439}, we introduce the superoperator
\begin{eqnarray}
    \mathcal{R}_\rho(X)\coloneqq\frac{1}{2}\left(\rho X+X\rho\right).
\end{eqnarray}
When expressed in the eigenbasis of $\rho$, this superoperator takes the form $\mathcal{R}_\rho(X)=\sum_{p_k+p_l>0}\frac{p_k+p_l}{2}X_{kl}\ket{k}\bra{l}$, where $X_{kl}=\bra{k}X\ket{l}$. It follows that the pseudoinverse of $\mathcal{R}_\rho$ is given by $\mathcal{R}_\rho^{-1}(X)=\sum_{p_k+p_l>0}\frac{2}{p_k+p_l}X_{kl}\ket{k}\bra{l}$ \cite{1994Braunstein3439} (see also Ref.~\cite{SM}). Letting $L\coloneqq\mathcal{R}_\rho^{-1}\left(i[\rho,H]\right)$, we obtain
\begin{eqnarray}
    F_Q=\norm{L}_\rho^2,
\end{eqnarray}
i.e., $F_Q$ can be compactly expressed as the squared norm of $L$.

{Second, we construct the Krylov subspaces with the aid of $\mathcal{R}_\rho$.}
Hereafter, given a set of Hermitian operators $\{X_k\}_{k=0}^{n-1}$, we denote the span of this set by $\text{span}\{X_k\}_{k=0}^{n-1}$, which consists of all linear combinations generated by the set, $\sum_{k=0}^{n-1}x_kX_k$, with real coefficients $x_k$. Let $\mathcal{R}_\rho^k$ denote the superoperator obtained by applying $\mathcal{R}_\rho$ iteratively $k$ times. By convention, $\mathcal{R}_\rho^0$ is the identity map. With these notations, we define the $n$th Krylov subspace as follows:
\begin{definition}
    $\mathcal{K}_n\coloneqq\text{span}\left\{\mathcal{R}_\rho^{k}(i[\rho,H])\right\}_{k=0}^{n-1}$.
\end{definition}
\noindent We refer to $\{\mathcal{R}_\rho^{k}(i[\rho,H])\}_{k=0}^{n-1}$ as the generating set of $\mathcal{K}_n$. Apparently, the generating set of $\mathcal{K}_{n+1}$ is constructed by augmenting that of $\mathcal{K}_n$ with $\mathcal{R}_\rho^n(i[\rho,H])$. Consequently, $\mathcal{K}_n$ expands as $n$ increases. This expansion, however, eventually terminates at a certain integer $n^*$, since the dimension of $\mathcal{K}_n$ cannot exceed that of the entire space of Hermitian operators. Therefore, we have
\textit{Observation 1:}
\begin{eqnarray}
    \mathcal{K}_1\subsetneq\mathcal{K}_2\subsetneq\cdots\subsetneq\mathcal{K}_{n^*}=\mathcal{K}_{n^*+j},
\end{eqnarray}
for any integer $j \geq 1$.
Moreover, we have
\textit{Observation 2:}
\begin{eqnarray}
    L\in\mathcal{K}_{n^*}~~\text{but}~~ L\notin\mathcal{K}_{n}~~ \text{whenever}~~ n<n^*.
\end{eqnarray}
The proofs of the two observations are presented in Ref.~\cite{SM}.
\begin{figure}
    \centering
    \includegraphics[width=0.9\linewidth]{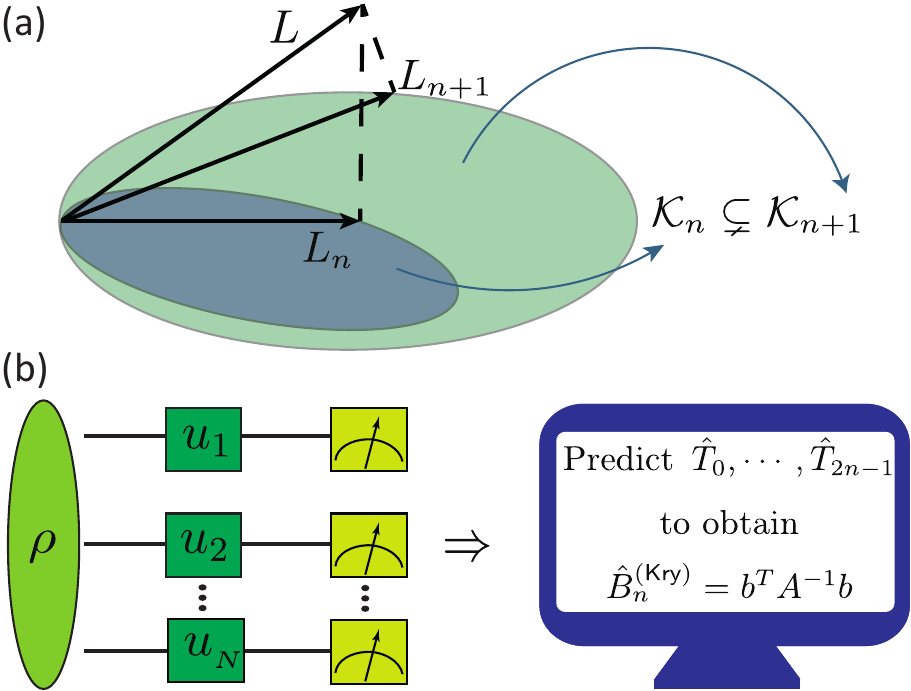}
    \caption{Schematic of the Krylov shadow tomography. (a) We construct a nested sequence of Krylov subspaces $\mathcal{K}_n$ and associate each of them with a non-polynomial lower bound $B_n^{(\mathsf{Kry})}=\norm{L_n}_\rho^2$, where $L_n\in\mathcal{K}_n$ is chosen to be as close to $L$ as possible. (b) We use shadow tomography to simultaneously predict $\hat{T}_0, \cdots, \hat{T}_{2n-1}$ and insert them into Eq.~(\ref{main-finding-2nd}) to obtain $\hat{B}_n^{(\mathsf{Kry})}$. Here, $u_i$ denotes a single-qubit gate selected at random \cite{SM}.
    }
    \label{fig1}
\end{figure}

{Third, we associate each $\mathcal{K}_n$ with a non-polynomial lower bound $B_n^{(\mathsf{Kry})}$.} To ensure that $B_n^{(\mathsf{Kry})}$ effectively approximates $F_Q=\norm{L}_\rho^2$, we define $B_n^{(\mathsf{Kry})}=\norm{L_n}_\rho^2$, where $L_n\in\mathcal{K}_n$ is chosen to be as close to $L$ as possible [see Fig.~\ref{fig1}(a)]. Formally,
\begin{definition}
    $B_n^{(\mathsf{Kry})}\coloneqq\norm{L_n}_\rho^2$ with $\norm{L-L_n}_\rho=\min_{X\in\mathcal{K}_n}\norm{L-X}_\rho$.
\end{definition}
\noindent To see that $B_n^{(\mathsf{Kry})}$ is indeed a lower bound, let us recall that $L_n$ attains $\min_{X\in\mathcal{K}_n}\norm{L-X}_\rho$ if and only if it satisfies
\begin{eqnarray}\label{orth-condition}
    \langle X, L-L_n\rangle_\rho=0,
\end{eqnarray}
for any $X\in\mathcal{K}_n$ \cite{Debnath}. Using Eq.~(\ref{orth-condition}) and noting that $L_n\in\mathcal{K}_n$, we have $\norm{L_n+L-L_n}_\rho^2=\norm{L_n}_\rho^2+\norm{L-L_n}_\rho^2$, i.e.,
\begin{eqnarray}\label{F-B-relation}
    F_Q=B_n^{(\mathsf{Kry})}+\norm{L-L_n}_\rho^2.
\end{eqnarray}
This implies $B_n^{(\mathsf{Kry})}\leq F_Q$. That $B_n^{(\mathsf{Kry})}$ is non-polynomial can be seen from its explicit expression presented below. Noting that $\norm{L-L_n}_\rho=\min_{X\in\mathcal{K}_n}\norm{L-X}_\rho\geq\min_{X\in\mathcal{K}_{n+1}}\norm{L-X}_\rho=\norm{L-L_{n+1}}_\rho$, we have $B_n^{(\mathsf{Kry})}\leq B_{n+1}^{(\mathsf{Kry})}$. Actually, we can show that $B_n^{(\mathsf{Kry})}< B_{n+1}^{(\mathsf{Kry})}$ if $n<n^*$ \cite{SM}, as $\mathcal{K}_n$ is a proper subset of $\mathcal{K}_{n+1}$ (see \textit{Observation 1}). Further, noting that $\norm{L-L_{n^*}}=0$ because $L\in\mathcal{K}_{n^*}$ (see \textit{Observation 2}), we deduce from Eq.~(\ref{F-B-relation}) that $B_{n^*}^{(\mathsf{Kry})}=F_Q$.

Now we can present the first part of our main finding as follows:
\begin{eqnarray}\label{main-finding-1st}
    B_1^{(\mathsf{Kry})}<B_2^{(\mathsf{Kry})}<\cdots<B_{n^*}^{(\mathsf{Kry})}=F_Q,
\end{eqnarray}
stating that $B_n^{(\mathsf{Kry})}$ is an increasingly better lower bound as $n$ increases and eventually matches the QFI exactly when $n=n^*$.

\textit{Estimation of our lower bounds via shadow tomography.---}We need to derive the explicit expression of $B_n^{\textsf{(Kry)}}$. Let
\begin{eqnarray}
    T_k\coloneqq\tr\left(i[\rho,H]\mathcal{R}_\rho^{k}(i[\rho,H])\right).
\end{eqnarray}
Interestingly, $T_k$, $k=0, 1, \cdots$, are polynomials of $\rho$ \cite{SM}. Specifically, the first two of them read $T_0=2\tr(\rho^2H^2-\rho H\rho H)$ and $T_1=\tr(\rho^3 H^2-\rho^2H\rho H)$ (see Ref.~\cite{SM} for the general expression of $T_k$).
We introduce
\begin{eqnarray}
    A=
    \begin{bmatrix}
        T_1    & T_2     & \cdots & T_n      \\
        T_2    & T_3     & \cdots & T_{n+1}  \\
        \vdots & \vdots  & \ddots & \vdots   \\
        T_n    & T_{n+1} & \cdots & T_{2n-1}
    \end{bmatrix},~~
    b=
    \begin{bmatrix}
        T_0    \\
        T_1    \\
        \vdots \\
        T_{n-1}
    \end{bmatrix}.
\end{eqnarray}
Notably, each ascending skew-diagonal of $A$ is constant, that is, $A$ is a Hankel matrix \cite{Fuhrman2012}. When $n\leq n^*$, $A$ is positive-definite and hence invertible \cite{SM}. With these notations, we can specify the rest part of our main finding as 
\begin{eqnarray}\label{main-finding-2nd}
    B_n^{(\mathsf{Kry})}=b^TA^{-1}b,
\end{eqnarray}
showing that the lower bounds $B_n^{(\mathsf{Kry})}$, albeit being non-polynomial in nature,  can be expressed in terms of the polynomials $T_k$ through $A$ and $b$ (see Ref.~\cite{SM} for the proof).

To see how to estimate $B_n^{(\mathsf{Kry})}$, let us recall that classical shadows can be efficiently generated by applying local randomized unitaries to $\rho$ followed by standard measurements in the computational basis [see Fig.~\ref{fig1}(b)]. A key advantage of classical shadows is their ability to predict multiple polynomials of $\rho$ simultaneously, provided a sufficient number of them are collected. This feature is particularly well-suited to our needs. Specifically, to estimate $B_n^{(\mathsf{Kry})}$, we generate a sufficient number of classical shadows and use them to predict $\hat{T}_0,\hat{T}_1,\cdots,\hat{T}_{2n-1}$ simultaneously. Here and throughout, quantities with hats denote their estimates. Inserting all these $\hat{T}_k$'s into Eq.~(\ref{main-finding-2nd}) yields an estimate of $B_n^{(\mathsf{Kry})}$. Details of this protocol are provided in Ref.~\cite{SM}. A thorough analysis \cite{SM} shows that the number of classical shadows required here scales similarly, up to a constant factor, to that needed in the work \cite{Rath2021PRL}, which represents the state of the art in estimating the QFI for large systems. Note that the work \cite{Rath2021PRL} aims to estimate $B_n^{(\mathsf{Tay})}$, which diverges significantly from the QFI when $n$ is small. Note also that, as shown below, $n^*$ is small in interesting contexts, for which $B_{n^*}^{(\mathsf{Kry})}=F_Q$. So, compared with the cutting-edge method \cite{Rath2021PRL}, our KST opens up the possibility of efficiently estimating the QFI itself without increasing resource demands.

\textit{Application 1: Pseudo-pure state.---}We consider the state
\begin{eqnarray}\label{pp-state}
    \rho_p=(1-p)\ket{\psi}\bra{\psi}+p\frac{\openone}{2^N},
\end{eqnarray}
where $0\leq p\leq 1$ and $\ket{\psi}$ is a pure state. The state (\ref{pp-state}), known as pseudo-pure state \cite{Gershenfeld1997S}, arises frequently in experiments \cite{Barbieri2003PRL,Saunders2010NP,Dai2014PRL} and can be prepared using various methods, such as temporal averaging \cite{Knill1998PRA}, spatial averaging \cite{Cory1998PD}, logical labeling \cite{Vandersypen1999PRL}, and cat-benchmark \cite{Knill2000N}. To quantitatively compare our bounds with the polynomial lower bounds proposed previously, we adopt the figure of merit, 
\begin{equation}\label{merit}
    \mathcal{E}=\frac{\abs{B-F_Q}}{F_Q},
\end{equation}
which characterizes the relative error of the bound $B$ in question.

\begin{figure}
    \centering
    \includegraphics[width=\linewidth]{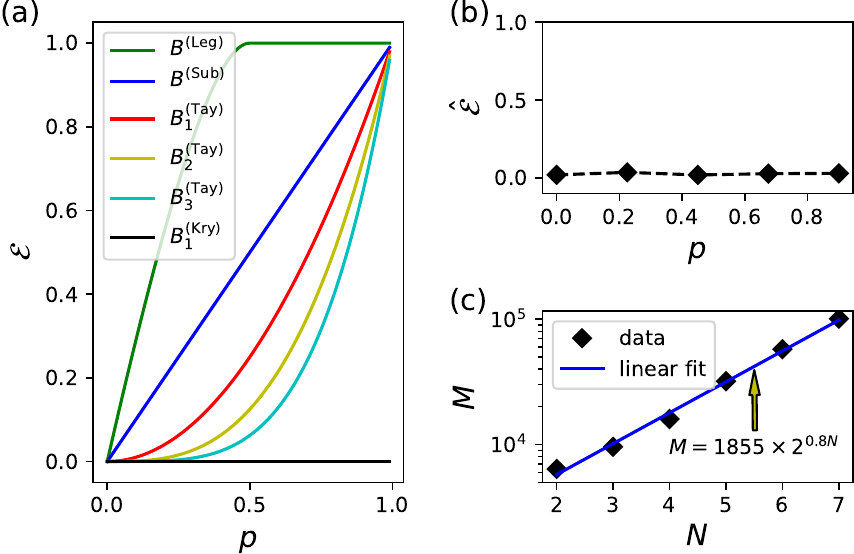}
    \caption{Illustrative results for application 1. We set $\ket{\psi}=\ket{GHZ}$ and $H=\frac{1}{2}\sum_{i=1}^N\sigma_z^{(i)}$. (a) $\mathcal{E}$ as a function of $p$ for different bounds with $N=12$. (b) $\hat{\mathcal{E}}$ as a function of $p$ for $B_1^{(\mathsf{Kry})}$ with $N=6$, predicted using the KST with $4.8\times 10^5$ classical shadows. (c) The required number $M$ of classical shadows for estimating $B_1^{(\mathsf{Kry})}$ with accuracy $\hat{\mathcal{E}}\leq 0.1$. Here $p=0.25$.
    }
    \label{fig2}
\end{figure}

Figure \ref{fig2}(a) shows $\mathcal{E}$ as a function of $p$ for different bounds. We see that the polynomial lower bounds deviate considerably from the QFI. This is expected since no polynomial can match the QFI exactly. Moreover, the mismatch becomes increasingly pronounced as $p$ increases. Notably, it was acknowledged as a difficulty in experiments to detect entanglement of the pseudo-pure state for relatively large $p$ \cite{Saunders2010NP}. While the QFI, if accessed in experiments, can tackle this difficulty \cite{Boixo2008PRL,Pezze2009PRL,Hyllus2012PRA,Toth2012PRA,Toth2018PRL,Toth2020PRL,Pal2021PRR,Ren2021PRL,Yadin2021NC,Tan2021PRL,Gianani2022PRA,Yang2024PRL,2024Imai,Szalay}, the polynomial lower bounds, unfortunately, may fail to do so due to the mismatch. In contrast, we show that $n^*=1$ and 
\begin{equation}\label{app1-match}
    B_1^{(\mathsf{Kry})}=F_Q,
\end{equation}
regardless of the value of $p$, the form of $\ket{\psi}$, and the choice of $H$ \cite{SM}. So, the KST can be used to fully unlock the effectiveness of the QFI in applications, as it provides an experimentally accessible bound that matches the QFI exactly.

To complement our analysis, we numerically simulate the KST. Figure \ref{fig2}(b) depicts $\hat{\mathcal{E}}$ as a function of $p$ for $B_1^{(\mathsf{Kry})}$ with $N=6$. Here, for each value of $p$, we use $4.8\times 10^5$ classical shadows to predict $\hat{B}_1^{(\mathsf{Kry})}$ and further obtain $\hat{\mathcal{E}}$ by inserting $\hat{B}_1^{(\mathsf{Kry})}$ into Eq.~(\ref{merit}). As can be seen from Fig.~\ref{fig2}(b), $\hat{B}_1^{(\mathsf{Kry})}$ matches $F_Q$ quite well. We further numerically study the scaling of the required number $M$ of classical shadows for estimating $B_1^{(\mathsf{Kry})}$ up to the relative error $0.1$. Figure \ref{fig2}(c) shows that $M$ scales as $2^{0.8N}$, which is the same as the resource demand for estimating $B_1^{(\mathsf{Tay})}$ \cite{Rath2021PRL}. This clearly demonstrates that the KST can efficiently estimate the QFI itself without increasing resource demands.

\textit{Application 2: Bound entangled state.---}We now consider the state
\begin{eqnarray}\label{bound-entangled-state}
    \rho_k=\lambda P_k^++\frac{\lambda}{2}(Q_k^++Q_k^-),
\end{eqnarray}
where $P_k^+=\sum_{\#1(i)<k}\ket{\phi_i^+}\bra{\phi_i^+}$ and $Q_k^{\pm}=\sum_{\#1(i)=k}\ket{\phi_i^\pm}\bra{\phi_i^\pm}$ \cite{Czekaj2015PRA}. Here, $1\leq k\leq\lfloor{N/2}\rfloor$, $\lambda=1/\sum_{i=0}^k\binom{N}{i}$, and $\ket{\phi_i^\pm}=\left(\ket{i}\pm\ket{\bar{i}}\right)/\sqrt{2}$, with $i=i_1i_2\cdots i_N$ in binary representation  and $\bar{i}=\bar{i}_1\bar{i}_2\cdots\bar{i}_N$. The notation $\#1(i)$ denotes the Hamming weight counting the number of ones in $i_1i_2\cdots i_N$. The state in Eq.~(\ref{bound-entangled-state}) is a bound entangled state \cite{Czekaj2015PRA}, whose entanglement has been acknowledged to be difficult to detect in experiments.

\begin{figure}
    \centering
    \includegraphics[width=\linewidth]{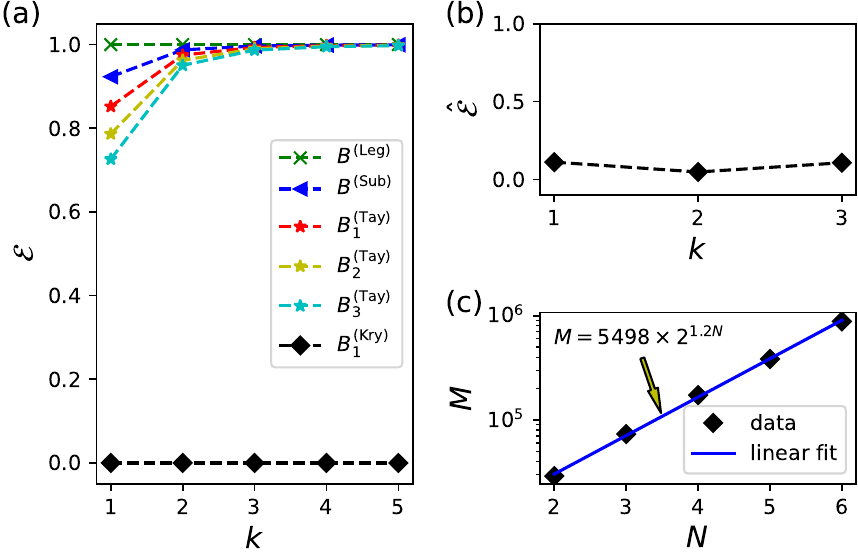}
    \caption{Illustrative results for application 2. We choose $H=\frac{1}{2}\sum_{i=1}^N\sigma_z^{(i)}$. (a) $\mathcal{E}$ as a function of $k$ for different bounds with $N=12$. (b) $\hat{\mathcal{E}}$ as a function of $k$ for $B_1^{(\mathsf{Kry})}$ with $N=6$, predicted using the KST with $3.2\times 10^6$ classical shadows. (c) The required number $M$ of classical shadows as a function of $N$ for estimating $B_1^{(\mathsf{Kry})}$ with accuracy $\hat{\mathcal{E}}\leq 0.1$. Here we set $k=1$.
    }
    \label{fig3}
\end{figure}

It has been shown that the QFI can effectively detect the bound entangled state when $H$ is chosen as $H=\frac{1}{2}\sum_{i=1}^N\sigma_z^{(i)}$ \cite{Czekaj2015PRA}. However, this effectiveness is severely compromised when the polynomial lower bounds are used as substitutes, owing to the significant gap between these bounds and the QFI [see Fig.~\ref{fig3}(a)]. In contrast, we show that $n^*=1$, implying that $B_1^{(\mathsf{Kry})}$ matches the QFI exactly \cite{SM}. So, we see again that the KST allows for fully unlocking the effectiveness of the QFI in applications.

To demonstrate the practical relevance of the match, we numerically simulate the KST to obtain an estimate $\hat{B}_1^{(\mathsf{Kry})}$ using $3.2\times 10^6$ classical shadows. As can be seen from Fig.~\ref{fig3}(b), $\hat{B}_1^{(\mathsf{Kry})}$ aligns closely with $F_Q$. Furthermore, Fig.~\ref{fig3}(c) illustrates that the required number of classical shadows, $M$, scales as $2^{1.2N}$. Notably, this scaling is more favorable compared with traditional methods like quantum state tomography, where the number of measurements scales as $\chi^2 2^N$ with $\chi$ denoting the rank of $\rho$ \cite{Haah2017IToIT}. 

\textit{Concluding remarks.}---We remark that $n^*$ is small if $\rho$ is of low rank, which arises frequently in practice \cite{Gross2010,Liu2012}. However, while $n^*$ is small in many scenarios, there are also cases where $n^*$ is large. In these cases, it is preferable to work in $\mathcal{K}_n$ with small $n$, as the current development in shadow tomography only allows for efficiently estimating polynomials of small order \cite{Zhou2024nQI,Peng-arXiv}. Notably, the QFI can usually be well approximated by $B_{n}^{(\mathsf{Kry})}$ with small $n$ due to the exponentially fast convergence of power iteration \cite{Saad2003}. So, even when $n^*$ is large, it is still expected that $B_{n}^{(\mathsf{Kry})}$'s with small $n$ are quite good lower bounds on the QFI. We present in Ref.~\cite{SM} some numerical results demonstrating the practical usefulness of $B_{n}^{(\mathsf{Kry})}$'s as lower bounds.

In conclusion, we have developed the KST to tackle the long-standing challenge of estimating the QFI for large systems. Our idea is to construct the nested sequence of Krylov subspaces $\mathcal{K}_n$ and associate each $\mathcal{K}_n$ with the non-polynomial lower bound $B_n^{(\mathsf{Kry})}$. We have shown that $B_n^{(\mathsf{Kry})}$ can approximate the QFI increasingly better as $n$ increases and eventually match it exactly when $\mathcal{K}_n$ stops expanding. This exact match is unattainable with polynomial lower bounds, which have been the focus of previous works. Further, we have shown that, despite their non-polynomial nature, all the bounds $B_n^{(\mathsf{Kry})}$ can be expressed in terms of the polynomials $T_k$ through $A$ and $b$, which enables $B_n^{(\mathsf{Kry})}$ to be accessible via shadow tomography.

We highlight that our KST allows for estimating not only non-polynomial lower bounds but, more importantly, the QFI itself in a resource-efficient and experimentally feasible manner. This offers a pathway to fully unlock the effectiveness of the QFI in vast applications. We have applied the KST to two interesting contexts, demonstrating that it opens the possibility of efficiently estimating the QFI itself without increasing resource demands.

Two directions for future works are to experimentally realize the KST and to further refine and extend it for estimating other physical quantities like various resource measures \cite{2009Horodecki865,2009Guehne1,2017Streltsov41003,2018Hu1,2018Zhang170501}.

%The codes and data used in this work are openly available from \cite{github}.

\begin{acknowledgments}
    This work was supported by the National Natural Science Foundation of China through Grants No.~12275155 and No.~12174224.
\end{acknowledgments}

\section*{Data Availability}
The data that support the findings of this article are openly available \cite{github}.

\renewcommand{\theequation}{S\arabic{equation}}
\setcounter{equation}{0}

\renewcommand{\thefigure}{S\arabic{figure}}
\setcounter{figure}{0}

\renewcommand{\thesection}{\Alph{section}}

\begin{center}
    \huge{\textbf{Supplemental Material}}
\end{center}

\section{Terminologies}
We begin by clarifying the terminologies used. Unless stated otherwise, all operators considered are assumed to be Hermitian. We denote the Hilbert-Schmidt inner product by
\begin{eqnarray}\label{HS-inner-product}
	(X, Y)\coloneqq\tr(X Y),
\end{eqnarray}
to distinguish it from the inner product used in the main text,
\begin{eqnarray}\label{weighted-inner-product}
	\langle X, Y\rangle_\rho\coloneqq\tr\left[\rho(XY+YX)/2\right].
\end{eqnarray}
To our knowledge, the inner product (\ref{weighted-inner-product}) was first introduced by Holevo \cite{2011Holevo}. Interestingly, the two inner products satisfy
\begin{eqnarray}\label{relation-inner-products}
	\langle X, Y\rangle_\rho=\left(\mathcal{R}_\rho(X), Y\right)=\left(X, \mathcal{R}_\rho(Y)\right),
\end{eqnarray}
where $\mathcal{R}_\rho(X)\coloneqq(\rho X+X\rho)/2$ is the superoperator introduced in the main text. Here we have used the fact that \textit{$\mathcal{R}_\rho$ is a Hermitian superoperator with respect to the Hilbert-Schmidt inner product}\footnote{A superoperator $\mathcal{S}$ is said to be Hermitian if $(\mathcal{S}(X), Y)=(X, \mathcal{S}(Y))$ for all $X$ and $Y$, analogous to the definition of Hermitian operators.}. Strictly speaking, Eq.~(\ref{weighted-inner-product}) only represents a pre-inner product when $\rho$ is singular, because it does not satisfy
the positive-definiteness requirement of an inner product. To resolve this issue, we introduce the subspace of Hermitian operators
\begin{eqnarray}
	\mathcal{X}\coloneqq\{X=X^\dagger: X_{kl}=0~~\text{if}~~p_k=p_l=0\},
\end{eqnarray}
where $p_k$ and $\ket{k}$ are the eigenvalues and eigenvectors of $\rho$, and $X_{kl}=\bra{k}X\ket{l}$, as specified in the main text.
The only potentially nonzero elements of $X\in\mathcal{X}$ are $X_{kl}$'s with $p_k+p_l>0$. Using the expression of $\mathcal{R}_\rho$ given in Ref.~\cite{1994Braunstein3439}
\begin{equation}\label{expression-R}
	\mathcal{R}_\rho(X)=\sum_{p_k+p_l>0}\frac{p_k+p_l}{2}X_{kl}\ket{k}\bra{l},
\end{equation}
we have that all the operators $\mathcal{R}_\rho^k(i[\rho, H])$, $k=0, 1, 2, \cdots$, belong to $\mathcal{X}$. Therefore, it suffices to restrict our analysis to the subspace $\mathcal{X}$. Hereafter, both the inner products, norms, and superoperators are assumed to be defined over $\mathcal{X}$. From Eq.~(\ref{expression-R}) and $\langle X, X\rangle_\rho=(\mathcal{R}_\rho(X),X)$, it follows that
\begin{eqnarray}
	\langle X, X\rangle_\rho=0~~\text{with}~~X\in\mathcal{X} ~~\text{if and only if}~~X=0,
\end{eqnarray}
that is, the positive-definiteness of Eq.~(\ref{weighted-inner-product}) is restored. This implies that Eq.~(\ref{weighted-inner-product}) is a legitimate inner product over $\mathcal{X}$. On the other hand, it is straightforward to verify that
\begin{eqnarray}
	\mathcal{R}_\rho(X)=0~~\text{with}~~X\in\mathcal{X} ~~\text{if and only if}~~X=0.
\end{eqnarray}
This implies that $\mathcal{R}_\rho$ is invertible over $\mathcal{X}$. We have denoted by $\mathcal{R}_\rho^{-1}$ the inverse of $\mathcal{R}_\rho$ in the main text. Explicitly,
\begin{eqnarray}
	\mathcal{R}_\rho^{-1}(X)=\sum_{p_k+p_l>0}\frac{2}{p_k+p_l}X_{kl}\ket{k}\bra{l},
\end{eqnarray}
which can be confirmed by verifying $\mathcal{R}_\rho^{-1}(\mathcal{R}_\rho(X))=X$ for any $X\in\mathcal{X}$ \cite{1994Braunstein3439}. Recall that the generating set of $\mathcal{K}_n$ reads
\begin{eqnarray}\label{generating-set}
	\left\{i[\rho,H], \mathcal{R}_\rho(i[\rho,H]), \cdots, \mathcal{R}_\rho^{n-1}(i[\rho,H])\right\}.
\end{eqnarray}
Throughout this work, we do not consider the trivial case that $[\rho, H]=0$, which implies that $F_Q=0$.
We say that the generating set satisfies \textit{the linear independence condition} if all its elements are linearly independent. In what follows, we strive to present the proofs of our results in a self-contained manner, though some of them may be simplified by referencing established theorems in the literature.

\section{Proofs of the two observations}

\begin{myobs}{}{obs1}
	The Krylov subspaces satisfy  $\mathcal{K}_1\subsetneq\mathcal{K}_2\subsetneq\cdots\subsetneq\mathcal{K}_{n^*}=\mathcal{K}_{n^*+1}=\mathcal{K}_{n^*+2}=\cdots$ for some positive integer $n^*$.
\end{myobs}

\begin{proof}
	Noting that the generating set of $\mathcal{K}_{2}$ is $\left\{i[\rho,H],\mathcal{R}_\rho(i[\rho,H])\right\}$ whereas that of $\mathcal{K}_1$ is $\left\{i[\rho,H]\right\}$, we have
	\begin{eqnarray}\label{inclusion-relation}
		\mathcal{K}_1\subset\mathcal{K}_{2}.
	\end{eqnarray}
	$\mathcal{K}_1$ may or may not be a proper subset of $\mathcal{K}_2$, depending on whether the generating set of $\mathcal{K}_{2}$ satisfies the linear independence condition. If the condition is satisfied, there is
	\begin{eqnarray}
		\text{dim}(\mathcal{K}_2)=2~~\text{whereas}~~\text{dim}(\mathcal{K}_1)=1,
	\end{eqnarray}
	where $\text{dim}(\mathcal{K}_n)$ denotes the dimension of $\mathcal{K}_n$.
	This implies that 
	\begin{eqnarray}
		\mathcal{K}_1\subsetneq\mathcal{K}_{2}.
	\end{eqnarray}
	Applying the above reasoning iteratively, we conclude that
	\begin{eqnarray}
		\mathcal{K}_n\subsetneq\mathcal{K}_{n+1},
	\end{eqnarray}
	provided that the generating set of $\mathcal{K}_{n+1}$ satisfies the linear independence condition. This means that $\mathcal{K}_n$
	expands as $n$ increases. However, this expansion must eventually terminate at some integer $n^*$, as the dimension of any Krylov subspace cannot exceed that of $\mathcal{X}$. The termination occurs when the generating set of $\mathcal{K}_{n^*}$ satisfies the linear independence condition whereas that of $\mathcal{K}_{n^*+1}$ does not. We have $\text{dim}(\mathcal{K}_{n^*})=\text{dim}(\mathcal{K}_{n^*+1})$, implying that
	\begin{eqnarray}\label{terminate-relation}
		\mathcal{K}_{n^*}=\mathcal{K}_{n^*+1}.
	\end{eqnarray}
	Note that the generating set of $\mathcal{K}_{n^*+1}$ is constructed by augmenting the generating set $\{\mathcal{R}_\rho^k(i[\rho, H])\}_{k=0}^{n^*-1}$ of $\mathcal{K}_{n^*}$ with $\mathcal{R}_\rho^{n^*}(i[\rho,H])$.
	From Eq.~(\ref{terminate-relation}), it follows that $\mathcal{R}_\rho^{n^*}(i[\rho, H])$ can be expressed as a linear combination of $\{\mathcal{R}_\rho^k(i[\rho, H])\}_{k=0}^{n^*-1}$. Using this fact iteratively, we have that $\mathcal{R}_\rho^{n^*+j}(i[\rho, H])$ is a linear combination of $\{\mathcal{R}_\rho^k(i[\rho, H])\}_{k=0}^{n^*-1}$ for any $j\geq 1$. This implies $\mathcal{K}_{n^*}=\mathcal{K}_{n^*+j}$ for $j\geq 1$.
\end{proof}

\begin{myobs}{}{obs2}
	The Hermitian operator $L\coloneqq\mathcal{R}_\rho^{-1}(i[\rho,H])$ belongs to the Krylov subspace $\mathcal{K}_{n^*}$ but not to $\mathcal{K}_{n}$ with $n<n^*$.
\end{myobs}

\begin{proof}
	We first prove $L\in\mathcal{K}_{n^*}$. Let $\mathcal{R}_\rho(\mathcal{K}_{n^*})$ denote the subspace
	\begin{eqnarray}\label{obs2-st1}
		\mathcal{R}_\rho(\mathcal{K}_{n^*})=\left\{\mathcal{R}_\rho(X): X\in\mathcal{K}_{n^*}\right\}.
	\end{eqnarray}
	Note that the generating set of $\mathcal{R}_\rho(\mathcal{K}_{n^*})$ is $\{\mathcal{R}_\rho^k(i[\rho,H])\}_{k=1}^{n^*}$, which is contained by the generating set of $\mathcal{K}_{n^*+1}$, i.e., $\{\mathcal{R}_\rho^k(i[\rho,H])\}_{k=0}^{n^*}$. It follows that
	\begin{eqnarray}\label{obs2-st2}
		\mathcal{R}_\rho(\mathcal{K}_{n^*})\subset\mathcal{K}_{n^*+1}.
	\end{eqnarray}
	Noting that $\mathcal{K}_{n^*}=\mathcal{K}_{n^*+1}$, we deduce from Eq.~(\ref{obs2-st2}) that
	\begin{eqnarray}\label{obs2-st3}
		\mathcal{R}_\rho(\mathcal{K}_{n^*})\subset\mathcal{K}_{n^*}.
	\end{eqnarray}
	Moreover, since $\mathcal{R}_\rho$ is invertible, the dimension of $\mathcal{R}_\rho(\mathcal{K}_{n^*})$ must be the same as that of $\mathcal{K}_{n^*}$. Hence, we obtain
	\begin{eqnarray}\label{obs2-st4}
		\mathcal{R}_\rho(\mathcal{K}_{n^*})=\mathcal{K}_{n^*}.
	\end{eqnarray}
	Note that $\mathcal{R}_\rho(L)=\mathcal{R}_\rho(\mathcal{R}_\rho^{-1}(i[\rho,H]))=i[\rho,H]$.
	Thus, $\mathcal{R}_\rho(L)\in\mathcal{K}_{n^*}$, which, in conjunction with Eq.~(\ref{obs2-st4}), leads to
	\begin{eqnarray}\label{obs2-st5}
		\mathcal{R}_\rho(L)\in\mathcal{R}_\rho(\mathcal{K}_{n^*}).
	\end{eqnarray}
	Using Eq.~(\ref{obs2-st5}) and the invertibility of $\mathcal{R}_\rho$, we obtain
	\begin{eqnarray}\label{obs2-st6}
		L\in\mathcal{K}_{n^*}.
	\end{eqnarray}
	To prove $L\notin\mathcal{K}_n$ with $n<n^*$, we assume, by contradiction, that
	\begin{eqnarray}\label{obs2-st7}
		L\in\mathcal{K}_n~~\text{with}~~n<n^*.
	\end{eqnarray}
	Then there is a linear combination of $\{\mathcal{R}_\rho^k(i[\rho,H])\}_{k=0}^{n-1}$ such that
	\begin{eqnarray}\label{obs2-st8}
		L=\sum_{k=0}^{n-1}x_k\mathcal{R}_\rho^k(i[\rho,H]),
	\end{eqnarray}
	where $x_k$ is a real coefficient. Using $\mathcal{R}_\rho(L)=i[\rho,H]$, we have
	\begin{eqnarray}\label{obs2-st9}
		i[\rho,H]=\sum_{k=1}^{n}x_k\mathcal{R}_\rho^k(i[\rho,H]),
	\end{eqnarray}
	meaning that $\mathcal{R}_\rho^k(i[\rho,H])$, $k=0,1,\cdots,n$, are linearly dependent.
	This contradicts the assumption that the generating set of $\mathcal{K}_{n+1}$ satisfies the linear independence condition.
\end{proof}

\section{Proof of the strict hierarchy of our bounds}\label{sec:strict-hierarchy}
\begin{mythm}{}{thm3}
	The non-polynomial lower bounds we formulate obey the strict hierarchy $B_1^{(\mathsf{Kry})}<B_2^{(\mathsf{Kry})}<\cdots <B_{n^*}^{(\mathsf{Kry})}$.
\end{mythm}

We have shown in the main text that $B_n^{(\mathsf{Kry})} \leq B_{n+1}^{(\mathsf{Kry})}$ for any $n\geq 1$. To establish that this inequality is strict for $n < n^*$, we need the assistance of the following two lemmas.

\begin{mylem}{}{lem4}
	There exit Hermitian operators $P_0$, $P_1$, $\cdots$, $P_{n^*-1}$ orthonormal with respect to the weighted inner product in Eq.~(\ref{weighted-inner-product}), i.e., $\langle P_k, P_l\rangle_\rho=\delta_{kl}$, such that
	$\mathcal{K}_n=\text{span}\{P_0,P_1,\cdots,P_{n-1}\}$, for $n=1, 2, \cdots, n^*$.
\end{mylem}

This lemma is a simple consequence of the Gram-Schmidt process for constructing an orthonormal basis from a set of vectors in an inner product space. As the proof is straightforward, we omit it. Lemma \ref{lem:lem4} provides a way to characterize the Krylov subspaces in terms of the orthonormal basis $\{P_0, P_1, \cdots, P_{n^*-1}\}$. The following lemma provides another way to characterize these subspaces.

\begin{mylem}{}{lem5}
	Let $Q_n=\mathcal{R}_\rho(L-L_n)$ with $n=0,1,\cdots,n^*-1$. Here, by convention, $L_0$ is set to be $0$. Then, there is $\mathcal{K}_n=\text{span}\{Q_0, Q_1, \cdots, Q_{n-1}\}$, for $n=1, 2, \cdots, n^*$.
\end{mylem}
\begin{proof}[Proof of Lemma \ref{lem:lem5}]
	We have pointed out in the main text that
	\begin{eqnarray}\label{lem5-st1}
		\langle X, L-L_n\rangle_\rho=0,
	\end{eqnarray}
	for any $X\in\mathcal{K}_n$. Using Eq.~(\ref{lem5-st1}) and the relation between the two inner products given by Eq.~(\ref{relation-inner-products}), we have that
	\begin{eqnarray}\label{lem5-st2}
		(X, Q_n)=(X, \mathcal{R}_\rho(L-L_n))=\langle X, L-L_n\rangle_\rho=0,
	\end{eqnarray}
	for any $X\in\mathcal{K}_n$. This implies that $Q_n$ is orthogonal to the Krylov subspace $\mathcal{K}_n$ with respect to the Hilbert-Schmidt inner product. Here, by saying ``$Q_n$ is orthogonal to $\mathcal{K}_n$,'' we mean $(Q_n, X)=0$ for all $X\in\mathcal{K}_n$. Note that
	\begin{eqnarray}\label{lem5-st3}
		\mathcal{R}_\rho(L_n)\in\mathcal{K}_{n+1},
	\end{eqnarray}
	which, in conjunction with $\mathcal{R}_\rho(L)=i[\rho,H]$, further leads to
	\begin{eqnarray}\label{lem5-st4}
		Q_n=i[\rho, H]-\mathcal{R}_\rho(L_n)\in\mathcal{K}_{n+1}.
	\end{eqnarray}
	As the dimension of $\mathcal{K}_{n+1}$ only increases by one in comparison with that of $\mathcal{K}_{n}$ when $n<n^*$, we therefore deduce that
	\begin{eqnarray}\label{lem5-st5}
		\mathcal{K}_{n+1}=\text{span}\{\mathcal{K}_n, Q_n\},
	\end{eqnarray}
	that is, the $(n+1)$th Krylov subspace is obtained by augmenting the $n$th Krylov subspace with $Q_n$. Here we have set $\mathcal{K}_0=0$ by convention. Using Eq.~(\ref{lem5-st5}) iteratively, we conclude that $\mathcal{K}_n=\text{span}\{Q_0, Q_1, \cdots, Q_{n-1}\}$ for $n\leq n^*$.
\end{proof}

Now, with Lemmas \ref{lem:lem4} and \ref{lem:lem5}, we are ready to prove Theorem \ref{thm:thm3}.

\begin{proof}[Proof of Theorem \ref{thm:thm3}]
	As stated in Lemma \ref{lem:lem4}, the orthonormal basis for $\mathcal{K}_{n^*}$ is $\{P_k\}_{k=0}^{n^*-1}$. Noting that $L\in\mathcal{K}_{n^*}$ (see Observation \ref{obs:obs2}), we can express $L$ in this basis as
	\begin{eqnarray}\label{thm3-st1}
		L=\sum_{k=0}^{n^*-1}\langle P_k, L\rangle_\rho P_k.
	\end{eqnarray}
	Likewise, since $L_n\in\mathcal{K}_n$ and the orthonormal basis for $\mathcal{K}_n$ is $\{P_k\}_{k=0}^{n-1}$ (see Lemma \ref{lem:lem4}), we can express $L_n$ in the form
	\begin{eqnarray}\label{thm3-st2}
		L_n=\sum_{k=0}^{n-1}x_k P_k,
	\end{eqnarray}
	where $x_k$'s are determined by minimizing $\norm{L-L_n}_\rho$ (see the main text for the definition of $L_n$). It is easy to see that
	\begin{eqnarray}\label{thm3-st3}
		x_k=\langle P_k, L\rangle_\rho.
	\end{eqnarray}
	Further, noting that $B_n^{(\mathsf{Kry})}=\norm{L_n}_\rho^2=\sum_{k=0}^{n-1}x_k^2$, we obtain
	\begin{eqnarray}\label{thm3-st4}
		B_n^{(\mathsf{Kry})}=\sum_{k=0}^{n-1}\langle P_k, L\rangle_\rho^2.
	\end{eqnarray}
	Likewise,
	\begin{eqnarray}
		B_{n+1}^{(\mathsf{Kry})}=\sum_{k=0}^{n}\langle P_k, L\rangle_\rho^2.
	\end{eqnarray}
	Hence, to show that $B_n^{(\mathsf{Kry})}<B_{n+1}^{(\mathsf{Kry})}$, it suffices to prove that
	\begin{eqnarray}\label{thm3-st5}
		\langle P_n, L\rangle_\rho\neq 0.
	\end{eqnarray}
	To this end, we assume, by contradiction, that $\langle P_n, L\rangle_\rho=0$. Using the relation between the two inner product, we have
	\begin{eqnarray}\label{thm3-st6}
		(P_n, Q_k)=(P_n, \mathcal{R}_\rho(L-L_k))=\langle P_n, L-L_k\rangle_\rho.
	\end{eqnarray}
	Besides, note that $P_n$ is orthogonal to $P_0$, $P_1$, $\cdots$, $P_{n-1}$ with respect to the weighted inner product. From Lemma \ref{lem:lem4}, it follows that $P_n$ is orthogonal to the Krylov subspace $\mathcal{K}_k$ with $k\leq n$. Noting that $L_k\in\mathcal{K}_k$, we have $\langle P_n, L_k\rangle_\rho=0$ for $k\leq n$.
	Then, from Eq.~(\ref{thm3-st6}), it follows that
	\begin{eqnarray}
		(P_n, Q_k)=\langle P_n, L\rangle_\rho,
	\end{eqnarray}
	which, under the assumption made, further implies that
	\begin{eqnarray}\label{thm3-st7}
		(P_n, Q_k)=0,
	\end{eqnarray}
	for all $k\leq n$. Given that $\{Q_k\}_{k=0}^{n}$ span $\mathcal{K}_{n+1}$ (see Lemma \ref{lem:lem5}), this implies that $P_n$ is orthogonal to $\mathcal{K}_{n+1}$ with respect to the Hilbert-Schmidt inner product. However, this contradicts the fact that $P_n\in\mathcal{K}_{n+1}$. Thus, the assumption is invalid, and we conclude that $\langle P_n, L\rangle_\rho\neq 0$.
\end{proof}

\section{General expressions of $T_k$}

Using $\mathcal{R}_\rho(X)=\frac{1}{2}(\rho X+X\rho)$, we have, after some algebra,
\begin{equation}\label{T-st1}
	\mathcal{R}_\rho^k(i[\rho,H])=\frac{1}{2^k}\sum_{l=0}^k\binom{k}{l}\rho^l\left(i[\rho,H]\right)\rho^{k-l}.
\end{equation}
Substituting Eq.~(\ref{T-st1}) into the defining expression of $T_k$
\begin{equation}
	T_k=\tr\left(i[\rho,H]\mathcal{R}_\rho^k(i[\rho,H])\right)
\end{equation}
and using the cyclic property of the trace, we have that 
\begin{widetext}
\begin{eqnarray}\label{T-st2}
	T_k=\frac{1}{2^k}\sum_{l=0}^{k}\binom{k}{l}\tr(H\rho^lH\rho^{k-l+2}-2H\rho^{l+1}H\rho^{k-l+1}
	+H\rho^{l+2}H\rho^{k-l}).
\end{eqnarray}
\end{widetext}
Then, reorganizing the terms in Eq.~(\ref{T-st2}) that involve $\tr(H\rho^{l}H\rho^{k-l+2})$, we arrive at the general expression of $T_k$,
\begin{equation}\label{expression-T}
	T_k=\frac{1}{2^k}\sum_{l=0}^{k+2} \mu_l^{(k)}\tr\left(H\rho^{l}H\rho^{k-l+2}\right),
\end{equation}
where 
\begin{equation}
	\mu_l^{(k)}=\binom{k}{l}-2\binom{k}{l-1}+\binom{k}{l-2}.
\end{equation}
Here, by convention, we set $\binom{k}{l}=0$ when $k<l$ or when $l<0$.

Further, for later usage, we rewrite Eq.~(\ref{expression-T}) in a different form. We introduce the cyclic permutation operator
\begin{equation}
	\Pi^{(k+2)}\ket{\psi_1}\otimes\ket{\psi_2}\otimes\cdots\otimes\ket{\psi_{k+2}}=
	\ket{\psi_2}\otimes\cdots\otimes\ket{\psi_{k+2}}\otimes\ket{\psi_{1}},
\end{equation}
where $\ket{\psi_j}$, $j=1,\cdots,k+2$, denote arbitrary pure states of the $N$ qubits. Then we can rewrite $\tr\left(H\rho^{l}H\rho^{k-l+2}\right)$ as 
\begin{eqnarray}\label{T-st3}
	&&\tr\left(H\rho^{l}H\rho^{k-l+2}\right)=\nonumber\\
	&&\tr[\Pi^{(k+2)}\left(H\otimes\openone^{\otimes(l-1)}\otimes H\otimes\openone^{\otimes(k-l+1)}\right)\rho^{\otimes(k+2)}].\nonumber\\
\end{eqnarray}
From Eq.~(\ref{T-st3}), it follows that $T_k$ can be expressed in the form
\begin{equation}\label{expression-T-alternative}
	T_k=\tr[O^{(k+2)}\rho^{\otimes(k+2)}],
\end{equation}
where
\begin{widetext}
\begin{eqnarray}\label{expression-O}
	O^{(k+2)}=\frac{1}{2^k}\sum_{l=0}^{k+2} \mu_l^{(k)}\Pi^{(k+2)}
	\left(H\otimes\openone^{\otimes(l-1)}\otimes H\otimes\openone^{\otimes(k-l+1)}\right)
\end{eqnarray}
\end{widetext}
is an operator acting on the $(k+2)$ copies of the $N$ qubits. Apparently,
\begin{equation}\label{permutation-rho}
	\pi\rho^{\otimes(k+2)}\pi^\dagger=\rho^{\otimes(k+2)},	
\end{equation}
where $\pi$ is a permutation operator for the $(k+2)$ copies of the $N$ qubits. Using Eq.~(\ref{permutation-rho}), we can rewrite Eq.~(\ref{expression-T-alternative}) as 
\begin{equation}\label{expression-T-alternative-scrambled}
	T_k=\tr[\overline{O}^{(k+2)}\rho^{\otimes(k+2)}].
\end{equation}
Here, we have introduced the ``symmetrized'' operator \cite{Rath2021PRL,Zhang2024PRL}
\begin{equation}\label{symmetrized-O}
	\overline{O}^{(k+2)}=\frac{1}{(k+2)!}\sum_{\pi}\pi^\dagger O^{(k+2)}\pi,
\end{equation}
where the sum is taken over all the $(k+2)!$ permutation operators $\pi$.

\section{Proof of the positive definiteness of $A$}
\begin{mypro}{}{thm6}
	The Hankel matrix $A$ is positive-definite when $n\leq n^*$.
\end{mypro}
\begin{proof}
	To maintain consistency with the notations in this work, we index the elements of
	$A$ using $k,l \in\{0, 1, \cdots, n-1\}$. That is, the $(k,l)$-element of $A$ is given by
	\begin{eqnarray}\label{thm6-st1}
		A_{kl}=T_{k+l+1},
	\end{eqnarray}
	which, in conjunction with the definition of $T_{k+l+1}$, leads to
	\begin{eqnarray}\label{thm6-st2}
		A_{kl}=\left(i[\rho,H],\mathcal{R}_\rho^{k+l+1}(i[\rho,H])\right).
	\end{eqnarray}
	As $\mathcal{R}_\rho$ and hence $\mathcal{R}_\rho^{k+l+1}$ are Hermitian superoperators with respect to the Hilbert-Schmidt inner product [see the text below Eq.~(\ref{relation-inner-products})],
	$A_{kl}$ is real, i.e., $A$ is a real matrix. To prove that $A$ is positive-definite, we need to show
	\begin{eqnarray}\label{thm6-st3}
		v^TAv> 0,
	\end{eqnarray}
	for any nonzero real vector $v=[v_0,v_1,\cdots,v_{n-1}]^T$. Note that
	\begin{eqnarray}\label{thm6-st4}
		\left(i[\rho,H],\mathcal{R}_\rho^{k+l+1}(i[\rho,H])\right)=\left\langle \mathcal{R}_\rho^{k}(i[\rho,H]),\mathcal{R}_\rho^{l}(i[\rho,H])\right\rangle_\rho,\nonumber
	\end{eqnarray}
	where we have used the Hermiticity of $\mathcal{R}_\rho$ with respect to the Hilbert-Schmidt inner product and the relation between the two inner product. From Eq.~(\ref{thm6-st2}), it follows that
	\begin{eqnarray}\label{expression-A-kl}
		A_{kl}=\langle \mathcal{R}_\rho^k(i[\rho,H]), \mathcal{R}_\rho^l(i[\rho,H])\rangle_\rho.
	\end{eqnarray}
	Then, after some simple algebra, we arrive at the equality,
	\begin{eqnarray}\label{thm6-st5}
		v^TAv=\norm{\sum_{k=0}^{n-1}v_k\mathcal{R}_\rho^{k}(i[\rho,H])}_\rho^2.
	\end{eqnarray}
	Besides, note that $\{\mathcal{R}_\rho^{k}(i[\rho,H])\}_{k=0}^{n-1}$ are linearly independent when $n\leq n^*$, which implies that $\sum_{k=0}^{n-1}v_k\mathcal{R}_\rho^{k}(i[\rho,H])\neq 0$. We deduce from Eq.~(\ref{thm6-st5}) that $v^TAv>0$.
\end{proof}

\section{Proof of the second part of our main finding}
\begin{mythm}{}{thm7}
	The bounds $B_n^{(\mathsf{Kry})}$ with $n\leq n^*$ can be expressed as
	\begin{eqnarray}\label{sm-main-finding-2nd}
		B_n^{(\mathsf{Kry})}=b^TA^{-1}b,
	\end{eqnarray}
	where $A$ is the $n\times n$ Hankel matrix with $A_{kl}=T_{k+l+1}$ and $b=[T_0,T_1,\cdots,T_{n-1}]^T$ is a real vector.
\end{mythm}

\begin{proof}
	Noting that $L_n\in\mathcal{K}_n$, we can express $L_n$ in the form
	\begin{eqnarray}\label{thm7-st1}
		L_n=\sum_{l=0}^{n-1}x_l\mathcal{R}_\rho^l(i[\rho,H]),
	\end{eqnarray}
	where $x_l$ is a real coefficient. Denote these coefficients collectively by $x=(x_0, x_1, \cdots, x_{n-1})^T$. Inserting Eq.~(\ref{thm7-st1}) into
	\begin{eqnarray}
		B_n^{(\mathsf{Kry})}=\norm{L_n}_\rho^2
	\end{eqnarray}
	and using Eq.~(\ref{expression-A-kl}),
	we obtain, after some simple algebra,
	\begin{eqnarray}\label{pre-expression-B}
		B_n^{(\mathsf{Kry})}=x^TAx.
	\end{eqnarray}
	It remains to determine $x$. To this end, recall that $L_n$ satisfies
	\begin{eqnarray}\label{thm7-st2}
		\langle X, L-L_n\rangle_\rho=0,
	\end{eqnarray}
	for any $X\in\mathcal{K}_n$. Noting that the generating set of $\mathcal{K}_n$ is $\{\mathcal{R}_\rho^k(i[\rho,H])\}_{k=0}^{n-1}$, we have that Eq.~(\ref{thm7-st2}) holds if and only if
	\begin{eqnarray}\label{thm7-st3}
		\langle \mathcal{R}_\rho^k(i[\rho,H]), L-L_n\rangle_\rho=0,
	\end{eqnarray}
	for all $k=0,1,\cdots,n-1$. We can rewrite Eq.~(\ref{thm7-st3}) as
	\begin{eqnarray}\label{thm7-st4}
		\langle \mathcal{R}_\rho^k(i[\rho,H]), L_n\rangle_\rho=\langle \mathcal{R}_\rho^k(i[\rho,H]), L\rangle_\rho.
	\end{eqnarray}
	Using Eq.~(\ref{thm7-st1}) and $A_{kl}=\langle \mathcal{R}_\rho^k(i[\rho,H]), \mathcal{R}_\rho^l(i[\rho,H])\rangle_\rho$, we have that the left-hand side of Eq.~(\ref{thm7-st4}) can be expressed as
	\begin{eqnarray}\label{thm7-st5}
		\langle \mathcal{R}_\rho^k(i[\rho,H]), L_n\rangle_\rho=\sum_{l=0}^{n-1}A_{kl}x_l.
	\end{eqnarray}
	On the other hand, noting that $\mathcal{R}_\rho(L)=i[\rho,H]$ and using the relation between the two inner product [see Eq.~(\ref{relation-inner-products})], we can express the right-hand side of Eq.~(\ref{thm7-st4}) as
	\begin{eqnarray}\label{thm7-st6}
		\langle \mathcal{R}_\rho^k(i[\rho,H]), L\rangle_\rho=T_{k}.
	\end{eqnarray}
	From Eqs.~(\ref{thm7-st5}) and (\ref{thm7-st6}), it follows that $x$ is determined by
	\begin{eqnarray}
		Ax=b.
	\end{eqnarray}
	Inserting $x=A^{-1}b$ into Eq.~(\ref{pre-expression-B}), we arrive at the expression (\ref{sm-main-finding-2nd}).
\end{proof}

\section{Details on the shadow tomography protocol}

\begin{figure}
    \centering
    \includegraphics[width=\linewidth]{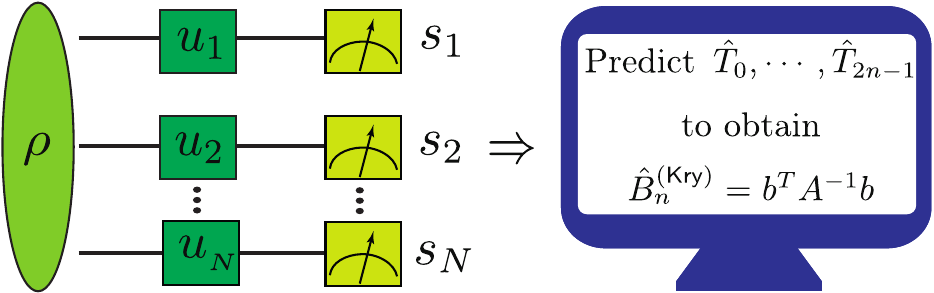}
    \caption{Schematic of the shadow tomography protocol for estimating $B_n^{(\mathsf{Kry})}$. Let $\rho$ be the unknown state of $N$ qubits. A randomly selected unitary operation $U=\bigotimes_{j=1}^Nu_j$ is applied to $\rho$, followed by the projective measurement in the computational basis with outcome recorded as a bit string $s=(s_1,s_2,\cdots,s_N)$. This procedure is repeated $M$ times. Consequently, $M$ classical shadows can be constructed, which can be employed to simultaneously predict $\hat{T}_0, \cdots, \hat{T}_{2n-1}$. Inserting these $\hat{T}_k$ into Eq.~(\ref{sm-main-finding-2nd}) finally produces an estimate $\hat{B}_n^{(\mathsf{Kry})}$.
    }
    \label{fig1-sm}
\end{figure}

We specify how to estimate $B_n^{(\mathsf{Kry})}$ using shadow tomography \cite{Elben2022NRP,Cieslinski2024PR}.
A schematic representation of the protocol is provided in Fig.~\ref{fig1-sm}. Let $U$ denote a unitary operation that is selected at random from a suitable ensemble of unitary operations.  Specifically, we consider the setting that
\begin{equation}\label{ST-U}
	U=\bigotimes_{j=1}^N u_j,
\end{equation}
where $u_j$, acting on qubit $j$, is drawn from an ensemble of single-qubit unitary operations that evenly cover the Bloch sphere. Examples of such ensembles include the single-qubit Clifford group and the full unitary group $U(2)$. We apply $U$ to the state $\rho$ and perform the measurement in the computational basis. The measurement outcome is recorded as a bit string $s=(s_1,s_2,\cdots,s_N)$. We repeat this procedure for $M$ randomly selected operations
\begin{equation}\label{ST-U}
	U^{(m)}=\bigotimes_{j=1}^N u_j^{(m)},
\end{equation}
with $m=1,2,\cdots,M$. Assigned to each $U^{(m)}$, a bit string 
\begin{equation}
	s^{(m)}=(s_1^{(m)},s_2^{(m)},\cdots,s_N^{(m)})
\end{equation}
is stored for recording the measurement outcome when $U^{(m)}$ is applied.
We associate each pair of $U^{(m)}$ and $s^{(m)}$ with 
\begin{equation}\label{shadow}
	\hat{\rho}^{(m)}=\bigotimes_{j=1}^N\left(3u_j^{(m)\dagger}\ket{s_j^{(m)}}\bra{s_j^{(m)}}u_j^{(m)}-\openone\right),
\end{equation}
where $\openone$ is the identity operator. The operator $\hat{\rho}^{(m)}$,
known as a classical shadow, reproduces the state $\rho$ exactly in expectation over both unitaries and measurement outcomes,
\begin{equation}
	\mathbb{E}[\hat{\rho}^{(m)}]=\rho.
\end{equation}
Intuitively speaking, $\hat{\rho}^{(m)}$ can be viewed as a classical snapshot of $\rho$. 
%Now, we have $M$ classical shadows at our disposal.

\begin{figure}
    \centering
    \includegraphics[width=\linewidth]{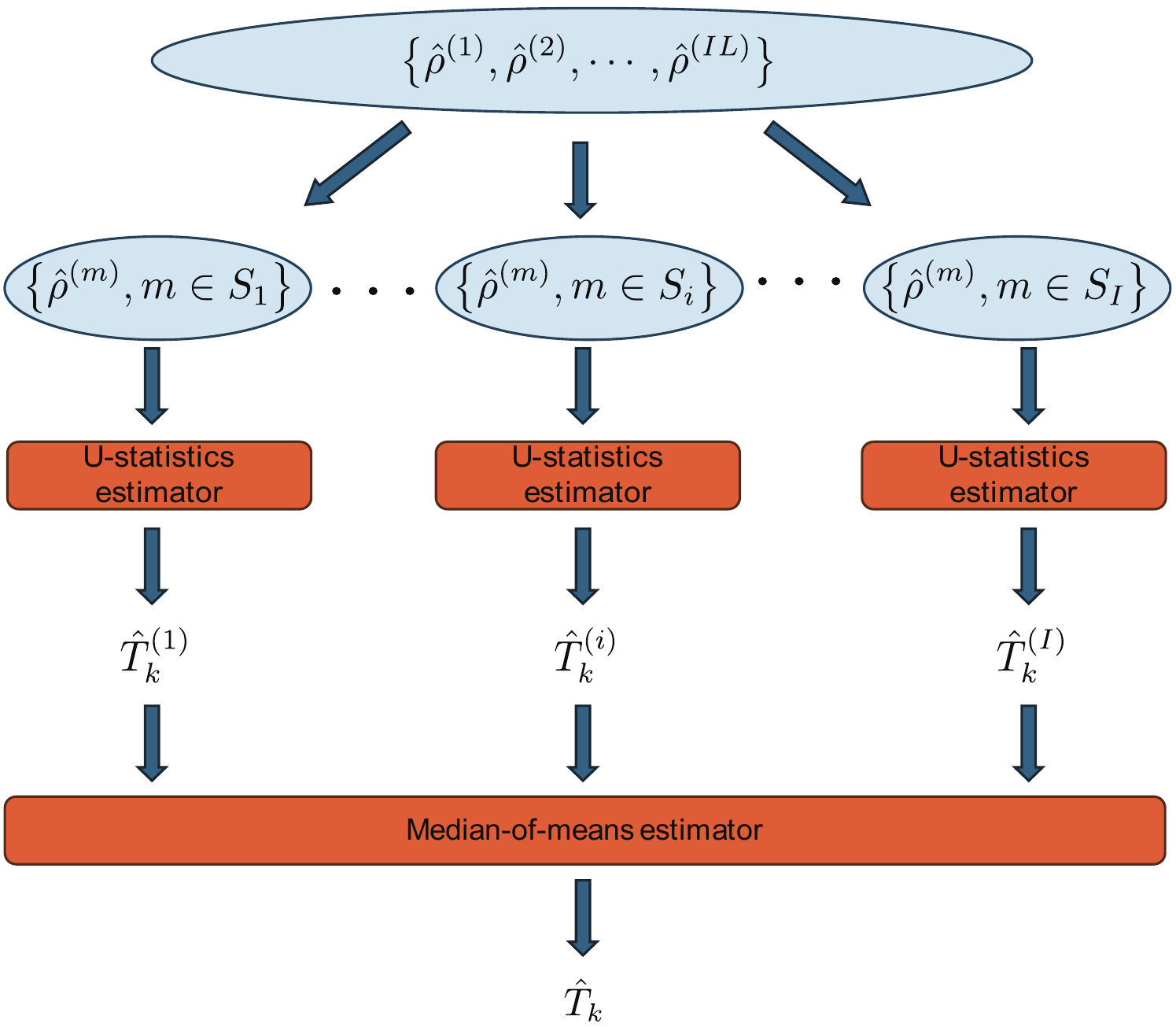}
    \caption{Flowchart of the procedure of classically postprocessing the classical shadows for obtaining $\hat{T}_k$. We divide the $M=IL$ classical shadows into $I$ equally sized subsamples and use each of the subsamples to obtain an estimate $\hat{T}_{k}^{(i)}$ via the $U$-statistics estimator. The final estimate $\hat{T}_k$ is produced from the estimates $\hat{T}_{k}^{(1)}, \hat{T}_{k}^{(2)}, \cdots, \hat{T}_{k}^{(I)}$ via the median-of-means estimator.
    }
    \label{fig2-sm}
\end{figure}

To estimate $B_n^{(\mathsf{Kry})}$, we need to use the $M$ classical shadows to predict $\hat{T}_0,\hat{T}_1,\cdots,\hat{T}_{2n-1}$ simultaneously \cite{Huang2020}, as clarified in the main text. Here and throughout this work, a quantity with a hat represents its estimate. For example, $\hat{B}_n^{(\mathsf{Kry})}$ is used to denote an estimate of $B_n^{(\mathsf{Kry})}$. We divide the $M$ classical shadows into $I$ equally sized subsamples. The $i$-th subsample consists of the classical shadows $\hat{\rho}^{(m)}$ with indexes $m$ belonging to the following set 
\begin{equation}
	S_i=\left\{(i-1)L+1,(i-1)L+2,\cdots,iL\right\},
\end{equation} 
where $L$ represents the size of each subsample. Accordingly, there are $M=IL$ classical shadows in total. Recall that 
\begin{equation}\label{expression-T-alternative-2}
	T_k=\tr[O^{(k+2)}\rho^{\otimes(k+2)}],
\end{equation}
i.e., $T_k$ is a $(k+2)$-order polynomial of $\rho$. The approach based on $U$-statistics \cite{Huang2020} can be used to estimate such polynomials,
\begin{widetext}
\begin{equation}\label{U-statistics}
	\hat{T}_k^{(i)}=\frac{1}{(k+2)!}\binom{L}{k+2}^{-1}\sum_{m_1\neq\cdots\neq m_{k+2}}\tr[O^{(k+2)}\bigotimes_{j=1}^{k+2}\hat{\rho}^{(m_j)}],
\end{equation}
\end{widetext}
where we require that all $m_j$, $j=1,2,\cdots,k+2$, belong to $S_i$, that is, all the classical shadows appearing in Eq.~(\ref{U-statistics}) are drawn from the $i$-th subsample. Using all the $I$ subsamples, we obtain a sequence of estimates of $T_k$, 
\begin{equation}
	\hat{T}_k^{(1)}, \hat{T}_k^{(2)}, \cdots, \hat{T}_k^{(I)}.
\end{equation}
To avoid the so-called outlier corruption \cite{Huang2020}, we adopt the median-of-means estimator to obtain a final estimate of $T_k$,  
\begin{equation}
	\hat{T}_k=\text{median}\left\{\hat{T}_k^{(1)}, \hat{T}_k^{(2)}, \cdots, \hat{T}_k^{(I)}\right\}.
\end{equation}
The above procedure for obtaining $\hat{T}_k$ is schematically shown in Fig.~\ref{fig2-sm}. Notably, all the $\hat{T}_k$, $k=0,1,\cdots,2n-1$, can be obtained simultaneously via this procedure. Finally, we insert these $\hat{T}_k$ in Eq.~(\ref{sm-main-finding-2nd}) to yield an estimate $\hat{B}_n^{(\mathsf{Kry})}$.

To finalize the description of the shadow tomography protocol, it remains to choose appropriate values for $I$ and $L$, and figure out the estimation error resulting from these choices. A detailed analysis in Sec.~\ref{sec:error} shows that the protocol works well when $I$ and $L$ are chosen as follows:
\begin{equation}\label{I}
	I=8\ln\frac{2n}{\delta},
\end{equation}
and
\begin{widetext}
\begin{equation}\label{L}
	L=\max_{1\leq l\leq k+2\leq 2n+1}\left[\left(\frac{4(k+2)(k+2)!^2\tr(\left[O_l^{(k+2)}\right]^2)}{l!(k-l+2)!^2\epsilon^2}\right)^{1/l}2^N+l-1\right],
\end{equation}
\end{widetext}
where $\epsilon$ and $\delta$ are prescribed accuracy parameters. The maximum in Eq.~(\ref{L}) is taken over all indices $k$ and $l$ satisfying $1\leq l\leq k+2\leq 2n+1$, and $O_l^{(k+2)}$ is defined as  
\begin{eqnarray}\label{O-l-k}
	O_l^{(k+2)}=\tr_{l+1,\cdots,k+2}\left[\overline{O}^{(k+2)}\left(\openone^{\otimes l}\otimes\rho^{\otimes(k-l+2)}\right)\right],\nonumber\\
\end{eqnarray}
where the partial trace is taken over the copies indexed by $l+1,l+2,\cdots,k+2$, and $\overline{O}^{(k+2)}$ is the symmetrized operator described by Eq.~(\ref{symmetrized-O}). Notably, $O_l^{(k+2)}$ is an operator acting on $l$ copies of the $N$ qubits. Formally, we summarize the result as a theorem (see Sec.~\ref{sec:error} for its proof):
\begin{mythm}{}{thm8}
Let $\epsilon$ and $\delta$ be prescribed accuracy parameters, and $I$ and $L$ be given by Eqs.~(\ref{I}) and (\ref{L}), respectively. Then, the estimate $\hat{B}_n^{(\mathsf{Kry})}$ obtained from the shadow tomography protocol obeys the inequality
\begin{equation}
	\abs{\hat{B}_n^{(\mathsf{Kry})}-B_n^{(\mathsf{Kry})}}<2n\epsilon\max_{0\leq k\leq 2n-1}\abs{\frac{\partial}{\partial T_k}B_n^{(\mathsf{Kry})}}
\end{equation}
with probability at least $1-\delta$. Here, $B_n^{(\mathsf{Kry})}$ is viewed as a function of $T_0, T_1, \cdots, T_{2n-1}$ through Eq.~(\ref{sm-main-finding-2nd}).
\end{mythm}

Theorem \ref{thm:thm8} shows that a collection of $IL$ randomly generated classical shadows enables accurate estimation of $B_n^{(\mathsf{Kry})}$ with high probability. In the main text, we demonstrate that $n^*$ is small in interesting contexts, for which $B_{n^*}^{(\mathsf{Kry})}=F_Q$. In these cases, our protocol can accurately estimate the quantum Fisher information (QFI) itself using the $IL$ classical shadows. Notably, $I$ is constant. Furthermore, the $L$ in Eq.~(\ref{L}) is proportional, up to a constant factor, to the number of classical shadows required for estimating the polynomial lower bound $B_n^{(\mathsf{Tay})}$ \cite{Rath2021PRL}. Note that Ref.~\cite{Rath2021PRL} represents the state of the art in estimating the QFI for large systems. Therefore, in comparison to the most advanced result obtained so far \cite{Rath2021PRL}, our KST opens up the possibility of efficiently estimating the QFI itself without increasing resource demands.

\section{Error analysis}\label{sec:error}

To prove Theorem \ref{thm:thm8}, we need the assistance of the following lemmas. 

\begin{mylem}{}{error-MoM}
    Let $I=8\ln\frac{2n}{\delta}$. Under the condition that $\text{Pr}\left(\abs{\hat{T}_k^{(i)}-T_k}\geq\epsilon\right)\leq\frac{1}{4}$ for all $k=0, 1, \cdots, 2n-1$, 
\begin{equation}\label{T-simultaneous-estimation}
	\text{Pr}\left(\bigcap_{k=0}^{2n-1}\left\{\abs{\hat{T}_k-T_k}<\epsilon\right\}\right)\geq 1-\delta,
\end{equation}
where $\text{Pr}(\cdot)$ is the probability of the event in question.
\end{mylem}

Lemma \ref{lem:error-MoM} provides a sufficient condition for accurately estimating all the $T_k$'s simultaneously with probability at least $1-\delta$. Notably, while $T_k$ is specified by Eq.~(\ref{expression-T}) in the present work, Lemma \ref{lem:error-MoM} holds as well if $T_k$ is used to represent any other physical property, e.g., the expectation value of an observable. This point can be easily seen from the proof presented below. Consequently, Lemma \ref{lem:error-MoM} generalizes one of the main results in Ref.~\cite{Huang2020}, which establishes a sufficient condition for accurately predicting multiple expectation values of observables simultaneously. 

\begin{proof}[Proof of Lemma \ref{lem:error-MoM}] 
Noting that the complement of the set $\bigcap_{k=0}^{2n-1}\left\{\abs{\hat{T}_k-T_k}<\epsilon\right\}$ is 
$\bigcup_{k=0}^{2n-1}\left\{\abs{\hat{T}_k-T_k}\geq\epsilon\right\}$, we have 
\begin{eqnarray}\label{lem9-st1}
	&&\text{Pr}\left(\bigcap_{k=0}^{2n-1}\left\{\abs{\hat{T}_k-T_k}<\epsilon\right\}\right)\nonumber\\
	&&=1-
	\text{Pr}\left(\bigcup_{k=0}^{2n-1}\left\{\abs{\hat{T}_k-T_k}\geq\epsilon\right\}\right).
\end{eqnarray}
Further, noting that $\text{Pr}\left(\bigcup_{k=0}^{2n-1}\left\{\abs{\hat{T}_k-T_k}\geq\epsilon\right\}\right)\leq \sum_{k=0}^{2n-1}\text{Pr}\left(\abs{\hat{T}_k-T_k}\geq\epsilon\right)$, we deduce from Eq.~(\ref{lem9-st1}) that 
\begin{equation}\label{lem9-st2}
	\text{Pr}\left(\bigcap_{k=0}^{2n-1}\left\{\abs{\hat{T}_k-T_k}<\epsilon\right\}\right) 
	\geq1-\sum_{k=0}^{2n-1}\text{Pr}\left(\abs{\hat{T}_k-T_k}\geq\epsilon\right).
\end{equation}
Therefore, to prove Eq.~(\ref{T-simultaneous-estimation}), it suffices to show that  
\begin{equation}\label{lem9-st3}
	\text{Pr}\left(\abs{\hat{T}_k-T_k}\geq\epsilon\right)\leq \frac{\delta}{2n},
\end{equation}
for all $k=0,1,\cdots,2n-1$. Recall that $\hat{T}_k$ is obtained from $\hat{T}_k^{(1)},\hat{T}_k^{(2)},\cdots,\hat{T}_k^{(I)}$ via the median-of-means estimator,
\begin{equation}\label{lem9-st4}
	\hat{T}_k=\text{median}\{\hat{T}_k^{(1)},\hat{T}_k^{(2)},\cdots,\hat{T}_k^{(I)}\}.
\end{equation}
Hence, the event $\left\{\abs{\hat{T}_k-T_k}\geq\epsilon\right\}$ occurs only when at least $I/2$ of the $\hat{T}_k^{(i)}$'s lie outside $\epsilon$-distance to $T_k$. That is,
\begin{equation}\label{lem9-st5}
	\left\{\abs{\hat{T}_k-T_k}\geq\epsilon\right\}\subset\left\{\sum_{i=1}^{I}\mathcal{I}\left(\abs{\hat{T}_k^{(i)}-T_k}\geq\epsilon\right)\geq\frac{I}{2}\right\},
\end{equation}
where $\mathcal{I}\left(\abs{\hat{T}_k^{(i)}-T_k}\geq\epsilon\right)$ denotes the indicator function of the event $\left\{\abs{\hat{T}_k^{(i)}-T_k}\geq\epsilon\right\}$, that is, this function outputs $1$ when $\abs{\hat{T}_k^{(i)}-T_k}\geq\epsilon$ and $0$ otherwise. We introduce
\begin{equation}\label{lem9-st6}
	Z_k^{(i)}=\mathcal{I}\left(\abs{\hat{T}_k^{(i)}-T_k}\geq\epsilon\right),
\end{equation}
which represents a random variable with the expectation value
\begin{equation}\label{lem9-st7}
	P_k\coloneqq\mathbb{E}[Z_k^{(i)}]=\text{Pr}\left(\abs{\hat{T}_k^{(i)}-T_k}\geq\epsilon\right).
\end{equation}
It is noteworthy that $\mathbb{E}[Z_k^{(i)}]$ is independent of $i$, as all the $I$ subsamples are identical from a statistical perspective. Using Eqs.~(\ref{lem9-st6}) and (\ref{lem9-st7}), we can rewrite Eq.~(\ref{lem9-st5}) as 
\begin{equation}\label{lem9-st8}
	\left\{\abs{\hat{T}_k-T_k}\geq\epsilon\right\}\subset\left\{\frac{1}{I}\sum_{i=1}^{I}\left(Z_k^{(i)}-\mathbb{E}[Z_k^{(i)}]\right)\geq\frac{1}{2}-P_k\right\}.
\end{equation}
Hence, 
\begin{eqnarray}\label{lem9-st9}
	&&\text{Pr}\left(\abs{\hat{T}_k-T_k}\geq\epsilon\right)\leq\nonumber\\
	&&\text{Pr}\left(\frac{1}{I}\sum_{i=1}^{I}\left(Z_k^{(i)}-\mathbb{E}[Z_k^{(i)}]\right)\geq\frac{1}{2}-P_k\right).
\end{eqnarray}
Using the one-side Hoeffding's inequality \cite{Mitzenmacher2005}, we can bound the probability in the right-hand side of Eq.~(\ref{lem9-st9}) as
\begin{equation}\label{lem9-st10}
	\text{Pr}\left(\frac{1}{I}\sum_{i=1}^{I}\left(Z_k^{(i)}-\mathbb{E}[Z_k^{(i)}]\right)\geq\frac{1}{2}-P_k\right)\leq e^{-2I(\frac{1}{2}-P_k)^2}.
\end{equation}
Substituting Eq.~(\ref{lem9-st10}) into Eq.~(\ref{lem9-st9}) and further noting that $P_k\leq\frac{1}{4}$ according to the condition in Lemma \ref{lem:error-MoM}, we have 
\begin{equation}\label{lem9-st11}
	\text{Pr}\left(\abs{\hat{T}_k-T_k}\geq\epsilon\right)\leq e^{-\frac{I}{8}}.
\end{equation}
Inserting $I=8\ln\frac{2n}{\delta}$ into Eq.~(\ref{lem9-st11}), we finally arrive at Eq.~(\ref{lem9-st3}).
\end{proof}

\begin{mylem}{}{var-T}
The variance of $\hat{T}_k^{(i)}$, denoted as $\text{Var}[\hat{T}_k^{(i)}]$, is bounded as
\begin{equation}\label{eq:var-T}
	\text{Var}[\hat{T}_k^{(i)}]\leq\sum_{l=1}^{k+2}\frac{(k+2)!^22^{lN}\tr\left(\left[O_l^{(k+2)}\right]^2\right)}{l!(k-l+2)!^2(L-l+1)^l},
\end{equation}
where $O_l^{(k+2)}$ is the $l$-copy operator given by Eq.~(\ref{O-l-k}).
\end{mylem}
Lemma \ref{lem:var-T} is credited to Rath \textit{et al.}~\cite{Rath2021PRL}.
Its proof can be found in Supplementary Material of Ref.~\cite{Rath2021PRL} [see Eq.~(D13)].

Now, with Lemmas \ref{lem:error-MoM} and \ref{lem:var-T}, we are ready to prove Theorem \ref{thm:thm8}.

\begin{proof}[Proof of Theorem \ref{thm:thm8}] 
	We first show how to choose an appropriate value for $L$ to meet the condition in Lemma \ref{lem:error-MoM},
	\begin{equation}\label{thm8-st1}
		\text{Pr}\left(\abs{\hat{T}_k^{(i)}-T_k}\geq\epsilon\right)\leq\frac{1}{4}.
	\end{equation}
	To do this, we resort to the Chebyshev's inequality \cite{Mitzenmacher2005} and obtain  
	\begin{equation}\label{thm8-st2}
		\text{Pr}\left(\abs{\hat{T}_k^{(i)}-T_k}\geq\epsilon\right)\leq\frac{\text{Var}[\hat{T}_k^{(i)}]}{\epsilon^2}.
	\end{equation}
	To ensure the validity of Eq.~(\ref{thm8-st1}), it suffices to require 
	\begin{equation}\label{thm8-st3}
		\text{Var}[\hat{T}_k^{(i)}]\leq\frac{\epsilon^2}{4}.
	\end{equation}
    From Lemma \ref{lem:var-T}, it follows that Eq.~(\ref{thm8-st3}) holds if each term in the right-hand side of Eq.~(\ref{eq:var-T}) is no larger than $\frac{\epsilon^2}{4(k+2)}$, i.e.,
	\begin{equation}\label{thm8-st4}
		\frac{(k+2)!^22^{lN}\tr\left(\left[O_l^{(k+2)}\right]^2\right)}{l!(k-l+2)!^2(L-l+1)^l}\leq\frac{\epsilon^2}{4(k+2)}.
	\end{equation}
	To guarantee that Eq.~(\ref{thm8-st4}) holds for any $k$ and $l$ satisfying $1\leq l\leq k+2$ and $k\leq 2n-1$, we can choose $L$ to be given by Eq.~(\ref{L}). Now, we apply Lemma \ref{lem:error-MoM} to show that  
	\begin{equation}\label{thm8-st5}
		\abs{\hat{B}_n^{(\mathsf{Kry})}-B_n^{(\mathsf{Kry})}}<2n\epsilon\max_{0\leq k\leq 2n-1}\abs{\frac{\partial}{\partial T_k}B_n^{(\mathsf{Kry})}},
	\end{equation}
	with probability at least $1-\delta$. Apparently, $B_n^{(\mathsf{Kry})}$ can be viewed as a function of $T_0,T_1,\cdots,T_{2n-1}$ [see Eq.~(\ref{sm-main-finding-2nd})], i.e.,
	\begin{equation}\label{thm8-st6}
		B_n^{(\mathsf{Kry})}=B_n^{(\mathsf{Kry})}(T_0,T_1,\cdots,T_{2n-1}).
	\end{equation}
	Accordingly, 
	\begin{equation}\label{thm8-st7}
		\hat{B}_n^{(\mathsf{Kry})}=B_n^{(\mathsf{Kry})}(\hat{T}_0,\hat{T}_1,\cdots,\hat{T}_{2n-1}),
	\end{equation}
	that is, $\hat{B}_n^{(\mathsf{Kry})}$ is obtained by substituting the estimates $\hat{T}_0,\hat{T}_1,\cdots,\hat{T}_{2n-1}$ into the function in Eq.~(\ref{thm8-st6}). Using the Taylor-series expansions, we obtain, up to the first order,
	\begin{equation}\label{thm8-st8}
		\hat{B}_n^{(\mathsf{Kry})}=B_n^{(\mathsf{Kry})}+\sum_{k=0}^{2n-1}\frac{\partial B_n^{(\mathsf{Kry})}}{\partial T_k}\delta T_k,
	\end{equation}
	where $B_n^{(\mathsf{Kry})}$ and $\frac{\partial B_n^{(\mathsf{Kry})}}{\partial T_k}$ are evaluated at the true values of $T_k$'s, and  
	\begin{equation}\label{thm8-st9}
		\delta T_k=\hat{T}_k-T_k.
	\end{equation}
	Using Lemma \ref{lem:error-MoM}, we know that, with the specified choices of $I$ and $L$, 
	\begin{equation}\label{thm8-st10}
		\abs{\delta T_k}\leq\epsilon,
	\end{equation}
	with probability at least $1-\delta$. Then, it follows immediately that Eq.~(\ref{thm8-st5}) also holds with probability at least $1-\delta$.
\end{proof}

\section{Proof of $n^*=1$ for the two applications}

Hereafter, for two operators $X$ and $Y$, we use $X\propto Y$ to represent $X=c Y$ for a coefficient $c$. We present a lemma that provides a necessary and sufficient condition for $n^*=1$.

\begin{mylem}{}{condition}
	The necessary and sufficient condition for $n^*=1$ is that 
	\begin{equation}\label{ns-condition}
		[\rho, H]\propto [\rho^2, H].	
	\end{equation}
\end{mylem}

\begin{proof}
	Recall that $n^*=1$ means $\mathcal{K}_2=\mathcal{K}_1$, that is, $\mathcal{K}_n$ stops growing at $n^*=1$. Note that 
	$\mathcal{K}_2=\mathcal{K}_1$ if and only if
	\begin{equation}\label{thm11-st1}
		\mathcal{R}_\rho\left(i[\rho,H]\right)\propto i[\rho,H].
	\end{equation}
	We have 
	\begin{equation}\label{thm11-st2}
		\mathcal{R}_\rho\left(i[\rho,H]\right)=\frac{i}{2}[\rho^2,H].
	\end{equation}
	The proof is completed by inserting Eq.~(\ref{thm11-st2}) into Eq.~(\ref{thm11-st1}).
\end{proof}

\begin{mypro}{}{}
	Let $\rho_p$ be the pseudo-pure state specified in the main text. Then
	\begin{equation}
		n^*=1,
	\end{equation}
	regardless of the value of $p$, the form of $\ket{\psi}$, and the choice of $H$.
\end{mypro}

\begin{proof}
Noting that $\rho_p=(1-p)\ket{\psi}\bra{\psi}+p\frac{\openone}{2^N}$, we easily have that 
\begin{equation}\label{thm11-st3}
	[\rho_p,H]=(1-p)\left[\ket{\psi}\bra{\psi}, H\right]
\end{equation}
and 
\begin{equation}\label{thm11-st4}
	[\rho_p^2,H]=\left((1-p)+\frac{p(1-p)}{2^{N-1}}\right)\left[\ket{\psi}\bra{\psi}, H\right].
\end{equation}
Using Eqs.~(\ref{thm11-st3}) and (\ref{thm11-st4}), we immediately arrive at Eq.~(\ref{ns-condition}).
\end{proof}

\begin{mypro}{}{}
Let $\rho_k$ be the bound entangled state specified in the main text. Then
\begin{equation}
	n^*=1,
\end{equation}
when $1\leq k\leq\lfloor{N/2}\rfloor$ and $H=\frac{1}{2}\sum_{i=1}^N\sigma_z^{(i)}$, with $\lfloor{\cdot}\rfloor$ the floor function.
\end{mypro}

\begin{proof}
	Let $\#0(i)$ and $\#1(i)$ be the number of zeros and ones in the binary representation $i=i_1i_2\cdots i_N$, respectively. Apparently,
	\begin{equation}\label{thm12-st1}
		H\ket{i}=\frac{1}{2}w_i\ket{i},
	\end{equation}
	and 
	\begin{equation}\label{thm12-st2}
		H\ket{\bar{i}}=-\frac{1}{2}w_i\ket{\bar{i}},
	\end{equation}
	where $w_i=\#0(i)-\#1(i)$ denotes the difference between $\#0(i)$ and $\#1(i)$. From Eqs.~(\ref{thm12-st1}) and (\ref{thm12-st2}), it follows that 
	\begin{equation}\label{thm12-st3}
		H\ket{\phi_i^+}=\frac{1}{2}w_i\ket{\phi_i^-},
	\end{equation}
	and 
	\begin{equation}\label{thm12-st4}
		H\ket{\phi_i^-}=\frac{1}{2}w_i\ket{\phi_i^+},
	\end{equation}
	where $\ket{\phi_i^\pm}=(\ket{i}\pm\ket{\bar{i}})/\sqrt{2}$, as specified in the main text. Then, using $Q_k^{\pm}=\sum_{\#1(i)=k}\ket{\phi_i^\pm}\bra{\phi_i^\pm}$, we easily have that 
	\begin{equation}\label{thm12-st5}
		\left[Q_k^++Q_k^-, H\right]=0.
	\end{equation}
	Noting that 
	\begin{eqnarray}\label{sm-bound-entangled-state}
		\rho_k=\lambda P_k^++\frac{\lambda}{2}(Q_k^++Q_k^-),
	\end{eqnarray}
	where $P_k^+=\sum_{\#1(i)<k}\ket{\phi_i^+}\bra{\phi_i^+}$, we deduce from Eq.~(\ref{thm12-st5}) that 
	\begin{equation}\label{thm12-st6}
		[\rho, H]=\lambda[P_k^+,H].
	\end{equation}
	Besides, it is not difficult to see that, when $1\leq k\leq\lfloor{N/2}\rfloor$, 
	\begin{equation}\label{thm12-st7}
		P_k^{+}P_k^{+}=P_k^{+},~~P_k^{+}Q_k^{+}=0,~~P_k^{+}Q_k^{-}=0.
	\end{equation} 
	Hence,
	\begin{equation}\label{thm12-st8}
		[\rho^2,H]=\lambda^2[P_k^+,H].
	\end{equation}
	Then, using (\ref{thm12-st6}) and (\ref{thm12-st8}), we easily arrive at Eq.~(\ref{ns-condition}). 
\end{proof}

\section{Numerical results on the usefulness of $B_n^{(\mathsf{Kry})}$ as lower bounds}

\begin{figure}
    \centering
    \includegraphics[width=\linewidth]{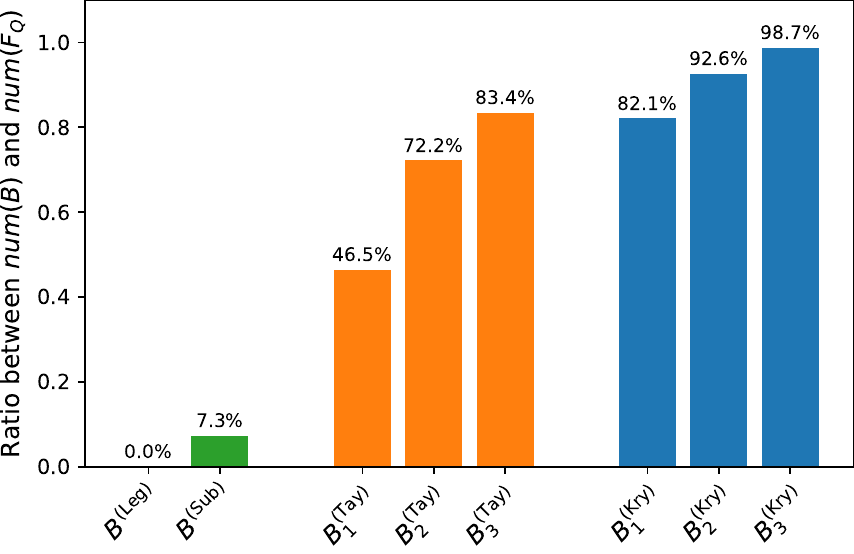}
    \caption{The ratio ${\textrm{num}(B)}\big/{\textrm{num}(F_Q)}$ for different lower bounds $B$. Here, ${\textrm{num}(F_Q)}$ denotes the number of states that is detected to be entangled by the QFI, and similarly, ${\textrm{num}(B)}$ represents the number of states that is detected to be entangled by the lower bound $B$ in question. The bars represent the values of the ratios for different bounds. The results are obtained by randomly generating ten million states for $N=2$ qubits.}
    \label{fig3-sm}
\end{figure}

To illustrate the usefulness of $B_n^{(\mathsf{Kry})}$ as a lower bound for the QFI, we explore its effectiveness in entanglement detection. To do this, we randomly generate ten million states $\rho$ for $N=2$ qubits and numerically figure out the ratio
\begin{equation}\label{sm:ratio}
	{\textrm{num}(B)}\big/{\textrm{num}(F_Q)}.
\end{equation}
Here, ${\textrm{num}(F_Q)}$ denotes the number of states that is detected to be entangled by the QFI; that is, $\rho$ is detected to be entangled if $F_Q>N$, where we choose $H=\frac{1}{2}\sum_{i=1}^N\sigma_z^{(i)}$. Similarly, ${\textrm{num}(B)}$ represents the number of states that is detected to be entangled by the lower bound $B$ in question, that is, $\rho$ is detected to be entangled if $B>N$. Physically speaking, the ratio described by Eq.~(\ref{sm:ratio}) quantifies the effectiveness of $B$ as a lower bound for the QFI in entanglement detection. Apparently, the closer the ratio is to $1$, the more effective the bound $B$ is. The numerical results are presented in Fig.~\ref{fig3-sm}. It is evident that $B^{(\mathsf{Leg})}$ and $B^{(\mathsf{Sub})}$ perform poorly as lower bounds, as they fail to detect entangled states effectively. By contrast, there are significant improvements in effectiveness for $B_n^{(\mathsf{Tay})}$ and $B_n^{(\mathsf{Kry})}$. Moreover, $B_n^{(\mathsf{Kry})}$ is significantly better than $B_n^{(\mathsf{Tay})}$ for the same $n$.
Note that, as already clarified, the resources required for estimating $B_n^{(\mathsf{Kry})}$ scale similarly to those required for estimating $B_n^{(\mathsf{Tay})}$ for the same $n$. Therefore, the results in Fig.~\ref{fig3-sm} demonstrate that $B_n^{(\mathsf{Kry})}$ can detect entanglement more effectively without increasing resource demands.

%\bibliography{refs}

\end{document}